\documentclass[journal,twoside,web]{ieeecolor}
\usepackage{tmi}

\usepackage{hyperref} 
\usepackage{times}
\usepackage{epsfig}
\usepackage{epsf}
\usepackage{pifont}
\usepackage{booktabs,multirow}
\usepackage{graphics}
\usepackage{threeparttable}
\usepackage{cite}
\usepackage{amsmath,amssymb,amsfonts}
\usepackage{textcomp}
\usepackage{threeparttable}
\usepackage{color}
\usepackage[normalem]{ulem}
\usepackage{multirow}
\usepackage{float}
\usepackage{amsfonts}
\usepackage{bbm}
\usepackage{multirow}
\usepackage{subfig}
\usepackage{array}
\usepackage[table]{xcolor}
\usepackage{colortbl}
\definecolor{bblue}{rgb}{0,150,230}
\definecolor{mygray}{gray}{.9}
\definecolor{myy}{RGB}{126,95,0}
\usepackage{bm}
\usepackage{bbding}
\usepackage{pifont}
\usepackage{wasysym}
\usepackage{amssymb}
\usepackage{graphicx} 
\usepackage{booktabs}

\definecolor{ggray}{RGB}{127,127,127}
\definecolor{mygreen}{RGB}{93,174,86}

\usepackage[ruled,linesnumbered]{algorithm2e}

\definecolor{myy}{RGB}{126,95,0}
\definecolor{mygray}{gray}{.9}
\definecolor{bblue}{RGB}{30,80,120}
\definecolor{mygray1}{gray}{.7}
\usepackage{mathtools}
\usepackage{siunitx}
\usepackage[nopar]{lipsum}
\usepackage[export]{adjustbox}
\usepackage{float}
\usepackage{amsfonts}
\usepackage{bm}
\usepackage[capitalize]{cleveref}
\crefname{section}{Sec.}{Secs.}
\Crefname{section}{Section}{Sections}
\Crefname{table}{Table}{Tables}
\crefname{table}{Tab.}{Tabs.}

\newcommand{\eg}[1]{\textit{e.g.,}}
\newcommand{\ie}[1]{\textit{i.e.,}}

\newcommand{\figref}[1]{Fig.\!~\ref{#1}}

\makeatletter
\newcommand{\thickhline}{%
	\noalign {\ifnum 0=`}\fi \hrule height 1pt
	\futurelet \reserved@a \@xhline
}
\makeatother

\usepackage{cleveref}
\crefname{section}{}{§§}
\Crefname{section}{}{§§}

\usepackage{caption}
\def\BibTeX{{\rm B\kern-.05em{\sc i\kern-.025em b}\kern-.08em
    T\kern-.1667em\lower.7ex\hbox{E}\kern-.125emX}}
    
\begin{document}

% \title{MTrans: Universal Transformer for Accelerated Multi-modal MR Imaging}
\title{Specificity-Preserving Federated Learning for MR Image Reconstruction}

\author{Chun-Mei Feng, Yunlu Yan,  Shanshan Wang, Yong Xu, \IEEEmembership{Senior Member, IEEE},\\  Ling Shao, \IEEEmembership{Fellow, IEEE}, 
and Huazhu Fu, \IEEEmembership{Senior Member, IEEE}% 
% and David Zhang, \IEEEmembership{Life Fellow, IEEE} 
\thanks{The work was supported by Shenzhen Science and Technology Innovation Committee: NO. GJHZ20210705141812038.}
\thanks{C.-M.~Feng is with the Shenzhen Key Laboratory of Visual Object Detection and Recognition, Harbin Institute of Technology (Shenzhen), 518055, China, and also with the Institute of High Performance Computing, A*STAR, Singapore 138632.~(Email: strawberry.feng0304@gmail.com)}
\thanks{Y.~Yan, and Y.~Xu are with the Shenzhen Key Laboratory of Visual Object Detection and Recognition, Harbin Institute of Technology (Shenzhen), 518055, China.~(Email: yongxu@ymail.com).}
\thanks{S.~Wang is with the Paul C. Lauterbur Research Center for Biomedical Imaging, Shenzhen Institutes of Advanced Technology, CAS, Shenzhen 518055, China. (Email: ss.wang@siat.ac.cn).}
\thanks{L.~Shao is with Terminus Group, China. (Email: ling.shao@ieee.org).}
\thanks{H.~Fu is with the Institute of High Performance Computing, A*STAR, Singapore 138632. (E-mail: hzfu@ieee.org).}
% \thanks{D. Zhang is with the School of Science and Engineering, The Chinese University of Hong Kong (Shenzhen), Shenzhen 518172, China, also with the Shenzhen Research Institute of Big Data, Shenzhen 518172, China, and also with the Shenzhen Institute of Artificial Intelligence and Robotics for Society, Shenzhen 518172, China (Email: davidzhang@cuhk.edu.cn)}National Center for Artificial Intelligence (NCAI), KSA
\thanks{Corresponding author: \textit{Yong Xu and Huazhu Fu}.}
\thanks{C.-M.~Feng and Y.~Yan are contributed equally to this work.}
}
\maketitle

\begin{abstract}

Federated learning (FL) can be used to improve data privacy and efficiency in magnetic resonance (MR) image reconstruction by enabling multiple institutions to collaborate without needing to aggregate local data. However, the domain shift caused by different MR imaging protocols can substantially degrade the performance of FL models. Recent FL techniques tend to solve this by enhancing the generalization of the global model, but they ignore the domain-specific features, which may contain important information about the device properties and be useful for local reconstruction. In this paper, we propose a specificity-preserving FL algorithm for MR image reconstruction (FedMRI). The core idea is to divide the MR reconstruction model into two parts: a globally shared encoder to obtain a generalized representation at the global level, and a client-specific decoder to preserve the domain-specific properties of each client, which is important for collaborative reconstruction when the clients have unique distribution.
Such scheme is then executed in the frequency space and the image space respectively, allowing exploration of generalized representation and client-specific properties simultaneously in different spaces. Moreover, to further boost the convergence of the globally shared encoder when a domain shift is present, a weighted contrastive regularization is introduced to directly correct any deviation between the client and server during optimization. Extensive experiments demonstrate that our FedMRI's reconstructed results are the closest to the ground-truth for multi-institutional data, and that it outperforms state-of-the-art FL methods.

\end{abstract}

\begin{IEEEkeywords}
MR image reconstruction, federated learning.
\end{IEEEkeywords}

\section{Introduction}\label{sec:intro}
\begin{figure}[t]
	\begin{center}
		\includegraphics[width=\linewidth]{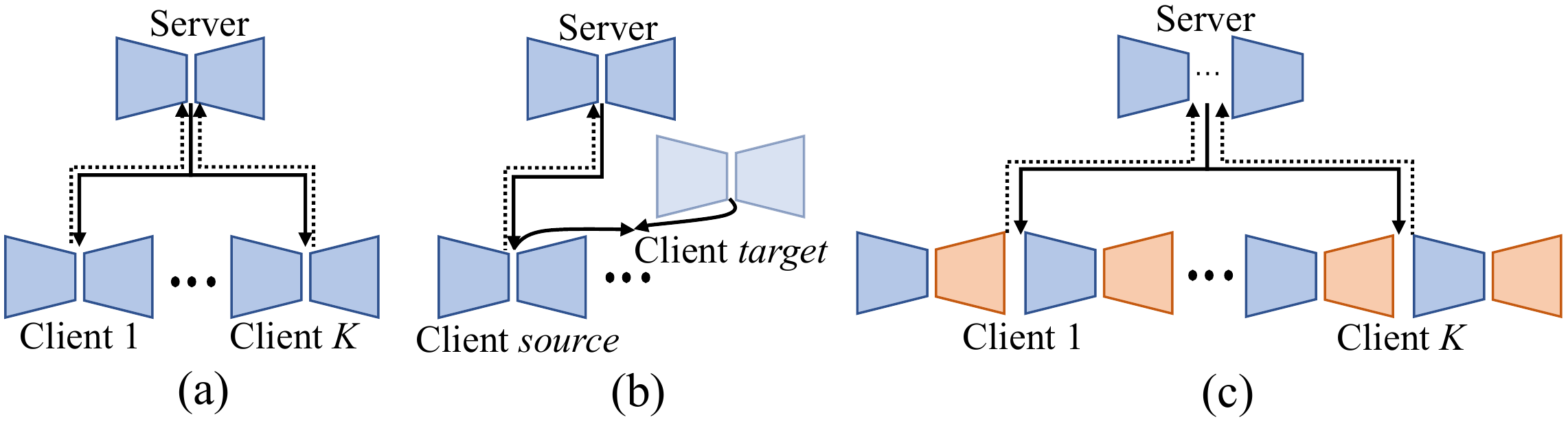}
% 		\put(-9,8){\tiny $K$} %(c) 
% 		\put(-173,8){\tiny $K$} %(a) 
	\end{center}
	\vspace{-8pt}
% 	\captionsetup{font=small}
	\caption{\!$_{\!}$Classical FL algorithm for MR image reconstruction: (a) average all the local client models to obtain a general global model~\cite{mcmahan2017communication}, or (b) repeatedly align the latent features between the source and target clients~\cite{guo2021multi}. In contrast, we propose a \textit{specificity-preserving} mechanism (c) to consider both ``generalized shared information'' as well as ``client-specific properties'' in both the frequency and image spaces.}
% 	\vspace{-13pt}
	\label{mo}
\end{figure}

% mainly due to its advantages of radiation-free, high spatial resolution and multi-contrast imaging.
Magnetic resonance (MR) imaging has become a mainstream diagnostic tool in radiology and medicine. However, its complex imaging process results in a longer acquisition time than other methods. To reduce the scanning time, several data-driven deep learning methods have been introduced, yielding outstanding improvements in MR image reconstruction~\cite{sriram2020grappanet,zhu2018image,zhou2020dudornet,sun2016deep,feng2022multi,feng2021donet,feng2021exploring}. However, the superior results obtained by deep learning-based methods are dependent on a large amount of diverse paired data, 
which is difficult to collect due to the patient privacy issues~\cite{Rieke2020}. Recently, federated learning (FL) was proposed to provide a platform for different clients to learn collaboratively using local computing power, memory, and data without sharing any private local data~\cite{mcmahan2017communication,Yang2019,RSCFed,FedDC}. One of the standard and most widely used FL algorithms, FedAvg~\cite{mcmahan2017communication}, collects local models of each client in each round of communication and distributes their average to each client for the next update (Fig.\!~\ref{mo}(a)). Privacy preservation is a major feature of FL approach, which allowing each client to collaborate training a global model without data sharing. However, the different MR scanners and imaging protocols of different hospitals present heterogeneity~\cite{guo2021multi}, leading to a domain shift between clients~\cite{quinonero2009dataset}. Unfortunately, the generalization of models trained using FL may still be suboptimal under these conditions~\cite{li2021fedbn,guo2021multi}. Guo \textit{et al.} tried to solve this problem by repeatedly aligning the latent features between the source and target clients (Fig.\!~\ref{mo}(b)), marking the first attempt to use FL in MR image reconstruction~\cite{guo2021multi}.

% % mainly due to its advantages of radiation-free, high spatial resolution and multi-contrast imaging.
% Magnetic resonance (MR) imaging has become a mainstream diagnostic tool in radiology and medicine. However, its complex imaging process results in a longer acquisition time than other methods, such as computed tomography (CT), X-ray, and ultrasound~\cite{sriram2020grappanet}. To reduce the scanning time and improve the patient experience, severl accelerated MR imaging methods have been proposed, \eg, the traditional methods of compressed sensing~\cite{otazo2010combination,liang2009accelerating}, dictionary learning~\cite{liu2020highly}, low ranking~\cite{haldar2016p,he2016accelerated} \etc. Recently, data-driven deep learning methods have also made outstanding improvements in MR image reconstruction, mainly thanks to the large amount of training data available~\cite{sriram2020grappanet,zhu2018image,zhou2020dudornet,hammernik2018learning}. However, the superior results obtained by deep learning-based methods are dependent on a large amount of diverse paired data, 
% which is difficult to collect due to the patient privacy issues~\cite{Rieke2020}.

% which is labor-intensive and difficult to collect because of patient privacy.  The Training data like this is hard to obtain, because health data is highly sensitive and its usage is tightly regulated.

% Recently, federated learning (FL) was proposed to provide a platform for different clients to learn collaboratively using local computing power, memory, and data without sharing any private local data~\cite{mcmahan2017communication,Yang2019}. One of the standard and most widely used FL algorithm, FedAvg~\cite{mcmahan2017communication}, collects local models of each client in each round of communication and distributes their average to each client for the next update. Benefiting from its use of distributed joint training, FL has been applied to many fields, including image classification~\cite{zhang2020federated,li2021model}, object detection~\cite{liu2020fedvision}, domain generalization~\cite{liu2021feddg}, medical image segmentation~\cite{li2019privacy,sheller2020federated}, etc. However, the different MR scanners and imaging protocols at different hospitals present heterogeneity~\cite{guo2021multi}, leading to a domain shift between clients~\cite{quinonero2009dataset}. Unfortunately, the generalization of models trained using FL may still be sub-optimal under these conditions~\cite{li2021fedbn,guo2021multi}. Guo \textit{et al.} tried to solve this problem by repeatedly aligning the latent features between the source and target clients, making the first attempt to use FL in MR image reconstruction~\cite{guo2021multi}.

Although FL has been applied to MR image reconstruction~\cite{guo2021multi}, this cross-site approach requires sacrificing one client as the target site for alignment with other clients in each round. It is obvious that any client being used as a target site will result in privacy concerns, and the cross-site approach contradicts the purpose of FL, which is to prevent clients from communicating with each other through local data. Further, it only trains each client in the image space, ignoring the properties of the client in the frequency space. When the number of clients is large, the process becomes tedious because of the repeated training and frequent feature communication. More importantly, such a mechanism can only learn a general global model, ignoring the \textit{client-specific properties}. Hence, we ask the following question: \textit{Does the server really need to average all the local client models during training?} Ideally, it should consider not only \textbf{1)} shared information but also \textbf{2)} client-specific properties in both image and frequency spaces. Prior studies~\cite{wang2018deep} in domain adaptation also suggest that the encoder is typically used to learn a shared representation to ensure that all inputs are equally suitable for any domain transformation. Therefore, we speculate that although FL algorithms have made some progress in MR image reconstruction~\cite{guo2021multi}, it is still possible to achieve higher-precision collaborative reconstruction when a domain shift is present by considering both \textbf{1)} and \textbf{2)}.

% Inspired by this,Could it instead share only part of the client model, such as the local encoders, leaving the remaining parts, such as the local decoders, to be kept local to learn the unique properties of each client?

% try to remove the domain shift among different clients without leakage of privacy from a new perspective: 
%

% Recently, Li \textit{et al.} used multi-level tasks framework to enhance the unsupervised representation learning, which proved that the low-level common representation is usually at the early of the network, while the deep information of different tasks is concentrated at the later stages~\cite{li2021progressive}. 

%如何在学习客户数据的共同表征的同时，保持客户数据的统计异质性
%How to retain the statistical heterogeneity of data among clients while learning the common representation of them?

With this insight, we propose a specificity-preserving FL algorithm for MR image reconstruction (FedMRI). Since deep neural networks can extract different information in different domains, we believe that the specific domain distribution of each client is not only in the image space but also in the frequency space. Here, we use two cascade U-Net~\cite{eo2018kiki} in a hybrid domain as the MR image reconstruction model and each is divided into two parts: a ``globally shared encoder'' to learn a \textit{generalized representation}, and a ``client-specific decoder'' to enable each client to explore their \textit{unique properties} under the presence of a domain shift (Fig.\!~\ref{mo}(c)). Further, to reduce the offset between client and server, we also introduce a weighted contrastive regularization term to \textit{correct} the update direction for the global generalization. Specifically, it pulls the current client model (\textit{anchor}) closer to the global model (\textit{positive}) and pushes it away from its last update (\textit{negative}). Our main \textbf{contributions} can be summarized as follows:
% \begin{itemize}
\begin{itemize} 
% 	\vspace{-5pt}
\item {We propose a \textit{specificity-preserving} FL algorithm for MR image reconstruction, FedMRI, which learns a generalized globally shared encoder in both the frequency space and the image space, while simultaneously providing a client-specific decoder for the local reconstruction. Our approach provides a more generalized representation and considers the local domain-specific information.}
\item {We develop a \textit{weighted contrastive regularization} for the globally shared part to relieve the domain shifts of clients during training and boost the convergence.}
% for further boosting the convergence of global-shared encoder in the domain shift setting, a weight contrastive regularization is introduced to directly corrects deviation between the client and server during optimization.
% boosting the convergence of global-shared encode
\item {We demonstrate that significant performance improvements can be achieved with our specificity-preserving mechanism and weighted contrastive regularization, compared to averaging the local client models. Moreover, we also show that our partially shared model reduces the communication costs.
\footnote{Our code is available at \url{https://github.com/chunmeifeng/FedMRI}.}}
\end{itemize}
% Experimental results show that our method provides significant performance improvements over several state-of-the-art FL methods on multi-source MR image datasets.

\section{Related Works}
% \vspace*{3pt} 
\noindent\textbf{MR Image Reconstruction.} 
Reconstructing an MR image from undersampled $k$-space data is an ill-posed inverse problem~\cite{sun2016deep,wang2020deepcomplexmri,zhang2020deep,zhang2017beyond,wang2016accelerating,feng2021dual,feng2021task}. Traditional methods typically solve the problem using prior knowledge~\cite{otazo2015low,majumdar2013non}. With the development of deep learning technology, CNN-based methods can directly learn the mapping from undersampled images to fully sampled data offline, which can greatly reduce the time of online reconstruction~\cite{feng2021multi}. Benefiting from research in natural image synthesis~\cite{zhang2021designing}, generative adversarial network (GAN)-based methods using a generator and discriminator with a loss function have also been used to reconstruct MR images. Recently, transformers have also been applied in MR image reconstruction, providing more global information than CNNs~\cite{korkmaz2022unsupervised,feng2021accelerated,feng2022multi}. However, all of these methods require the collection of large amounts of data. For medical data, this is not only labor-intensive but can also compromise patient privacy. We combine the data from different institutions for distributed joint training through FL, avoiding the dependence on large-scale data and the leakage of patients' privacy.

% \vspace*{3pt} 
\noindent\textbf{Federated Learning.}  
FL enables different clients to collaborate in training shared models, while maintaining data privacy~\cite{mcmahan2017communication}. However, the classical FL algorithm, FedAvg, does not address the domain shift caused by the statistical heterogeneity across various clients. More recent studies solve the feature shift and domain generalization problem caused by different data distributions through batch normalization and episodic learning in a continuous frequency space~\cite{li2021fedbn,liu2021feddg}. In the classification task, efforts have also been made to keep parts of the network local to improve the accuracy~\cite{arivazhagan2019federated,collins2021exploiting,liang2020think}. For example, to provide a global head, Liang \textit{et al.} sends the classification head to the server for averaging~\cite{liang2020think}, which is the opposite of our approach. Arivazhagan \textit{et al.} only explores one personalized layer after the base part~\cite{arivazhagan2019federated}. However, the data typically shares a global representation in the body, and the statistical heterogeneity between clients or tasks is concentrated in the head. Their model must therefore first learn a generalized representation, and then use the local head to train the client-specific data on the client.

FL has also been applied to some medical tasks to protect patient privacy~\cite{li2019privacy,roth2020federated,li2020multi,xue2022robust}. For example, Liu \textit{et al.} solved the problem of insufficient client labeling in the federal scenario by establishing disease relationships between labeled and unlabeled cases~\cite{liu2021federated}. Jiang \textit{et al.} tried to address the non-IID issue in medical images from the perspective of solving two essentially coupled drifts~\cite{jiang2021harmofl}. Dayan \textit{et al.} solved the joint training problem of COVID-19 data of medical institutions in different countries through federated learning~\cite{dayan2021federated}. Zheng \textit{et al.} introduced the concepts of local and global popularity and constructed an unsupervised recurrent federated learning (URFL) algorithm to predict popularity while achieving privacy-preserving goals~\cite{zheng2021privacypreserving}. There have also been several works developed that focus on collaborative training of COVID-19 data in different countries\cite{yang2021federated,dayan2021federated}. Many researchers are increasingly interested in the use of FL mechanisms to protect the privacy of healthcare data~\cite{kaissis2021end,usynin2021adversarial,sadilek2021privacy}. In MR image reconstruction, Guo \textit{et al.} repeatedly aligned the data distribution between the source and target client to reduce the data heterogeneity between them~\cite{guo2021multi}. However, this frequent communication between clients may reduce the communication efficiency and cause privacy leakage. Therefore, the effectiveness of the privacy protection mechanism of MR image reconstruction based on FL needs to be further studied. We introduce a \textit{specificity-preserving} FL mechanism that considers both the shared information and unique properties of clients in both frequency space and image spaces. Though a few concurrent works also consider using a personalized layer for classification tasks, their premise is markedly different~\cite{arivazhagan2019federated,liang2020think}. First, they often assume that only the last layer can be considered as the personalized part. Second, they reduce the domain shift problems by learning global heads, which is opposite to our approach (see \S\ref{sec:exp} for more details). Third, no one has ever attempted to progressively correct the update direction during training to reduce offsets and speed up convergence (\S\ref{sec:we}).

\noindent\textbf{Shared Representation Learning.}   
In domain adaptation, the encoder is often used to learn a shared representation to ensure that all inputs are equally suitable for any domain transformation~\cite{hu2018duplex,wang2018deep,cicek2019unsupervised}. This has also been applied to other fields~\cite{sanchez2020learning,nam2016learning,hu2018duplex}. For example, for classification, Sanchez \textit{et al.}~\cite{sanchez2020learning} used a low-dimensional shared representation to capture a generalized representation between images. Nam \textit{et al.}~\cite{nam2016learning} used shared layers to learn the generalized features of different domains, while employing the domain-specific layers of different branches to learn domain-specific properties for visual tracking. This inspired us to solve the domain shift problem of multi-institutional MR image reconstruction with a universal feature representation. Differently, however, we focus on privacy-protected distribution computing, where no data is communicated between clients.

% This is similar to our approach for solving the domain shift problem of multi-institutional MR image reconstruction with a universal feature representation. 

\section{Proposed Method}
\subsection{FL for MR Image Reconstruction}
MR image reconstruction aims to directly learn a mapping from undersampled to fully-sampled data in the image domain, where the undersampled data is obtained by accelerating the $k$-space data acquisition in the frequency space~\cite{qin2018convolutional,feng2021dual,guo2021multi}. An image can then be obtained by inverse multi-dimensional Fourier transform $\mathcal{F}^{-1}$ of the measured $k$-space points. Let $\mathbf{k}\in\mathbb{C}^{M}$ represent the observed $k$-space measurement, and $\mathbf{x}\in\mathbb{C}^{M}$ represent the undersampled image. The relationship between them can be formulated as follows:
\vskip -10pt
\begin{equation} 
\mathbf{x}=\mathcal{F}^{-1}(\mathbf{k}+\epsilon)=\mathcal{F}^{-1}\left(M \odot \mathcal{F}(\mathbf{y})+\epsilon\right),
\label{eq1}
\end{equation}
where $\epsilon$ is the measurement noise, $\odot$ is an element-wise multiplication operation, $M$ is the binary mask operator to select a subset of the $k$-space points, and $\mathbf{y}$ is the fully sampled image. Reconstructing $\mathbf{x}$ from the limited sampled data is an ill-posed inverse problem and we can obtain a more feasible solution by CNNs:
\vskip -10pt
\begin{equation}
\mathcal{L}_{rec}=\frac{1}{N} \sum_{n=1}^{N}\left\|f\left(\mathbf{x};\boldsymbol{w}\right)-\mathbf{y}\right\|_{1},
\label{eq:2}
\end{equation}
where $f\left(\mathbf{x}\right)$ is the network parameterized by $\boldsymbol{w}$, and $N$ is the number of training samples. 
Eq.~\eqref{eq:2} trains all of the hospital data together, resulting in the privacy of patients being leaked. If we train the data of each hospital in a distributed way using an FL framework, patient privacy can be protected. Suppose there are $S$ hospitals/clients. The local datasets of each client can be denoted as $\mathcal{D}^{1}, \mathcal{D}^{2}, \ldots, \mathcal{D}^{S}$, which contain pairs of undersampled and fully-sampled images. For each client, we train a local model $G^s$ by minimizing the following loss:
\vskip -10pt
\begin{equation}
\mathcal{L}_{rec}= \frac{1}{S}\sum_{s=1}^{S}\left\| G^s\left(\mathbf{x};\Theta_{G^{s}}\right)-\mathbf{y}\right\|_{1},
\label{eq:3}
\end{equation}
where $G^s\left(\mathbf{x};\Theta_{G^{s}}\right)$ corresponds to the reconstructed image $\hat{\mathbf{y}}$, and $G^s$ is the local model of client $s$ parameterized by $\Theta_{G^{s}}$. In each round $z=1,...,Z$, the local model parameter of client $s$ at local epoch $t$ is updated by the following optimization:
\vskip -10pt
\begin{equation}
\Theta_{G^{s}}^{z,t+1} = \Theta_{G^{s}}^{z,t}-\eta_{s} \nabla \mathcal{L}\left(\mathbf{x}^{s};\Theta_{G^{s}}^{z,t}\right),
\end{equation}
where $\eta_{s}$ is the learning rate of client $s$, and $t$ is the number of local updates, $\mathcal{L}$ is the total loss. All participating clients send their updated weights to the central server after completing each round of training locally. The central server then provides a global model by averaging all local parameters. It can be defined as follows:
\vskip -10pt
\begin{equation}
\Theta_{G}^{z}=\frac{1}{S} \sum_{s=1}^S \Theta_{G^{s}}^{z},
\end{equation}
where $\Theta_{G^{s}}^{z}$ represents the model parameter of client $s$ in the current global training round $z$. After $Z$ rounds of communication, the final global model parameterized by $\Theta_{G}^{Z}$ is trained with multi-domain information without directly sharing the private data of each client.

%%%%%%%%%%%%%%%%%%%%%%%%%%%%%%%%%%

\subsection{Architecture of FedMRI}
\begin{figure}[!t]
\centering
  \includegraphics[width=1\linewidth]{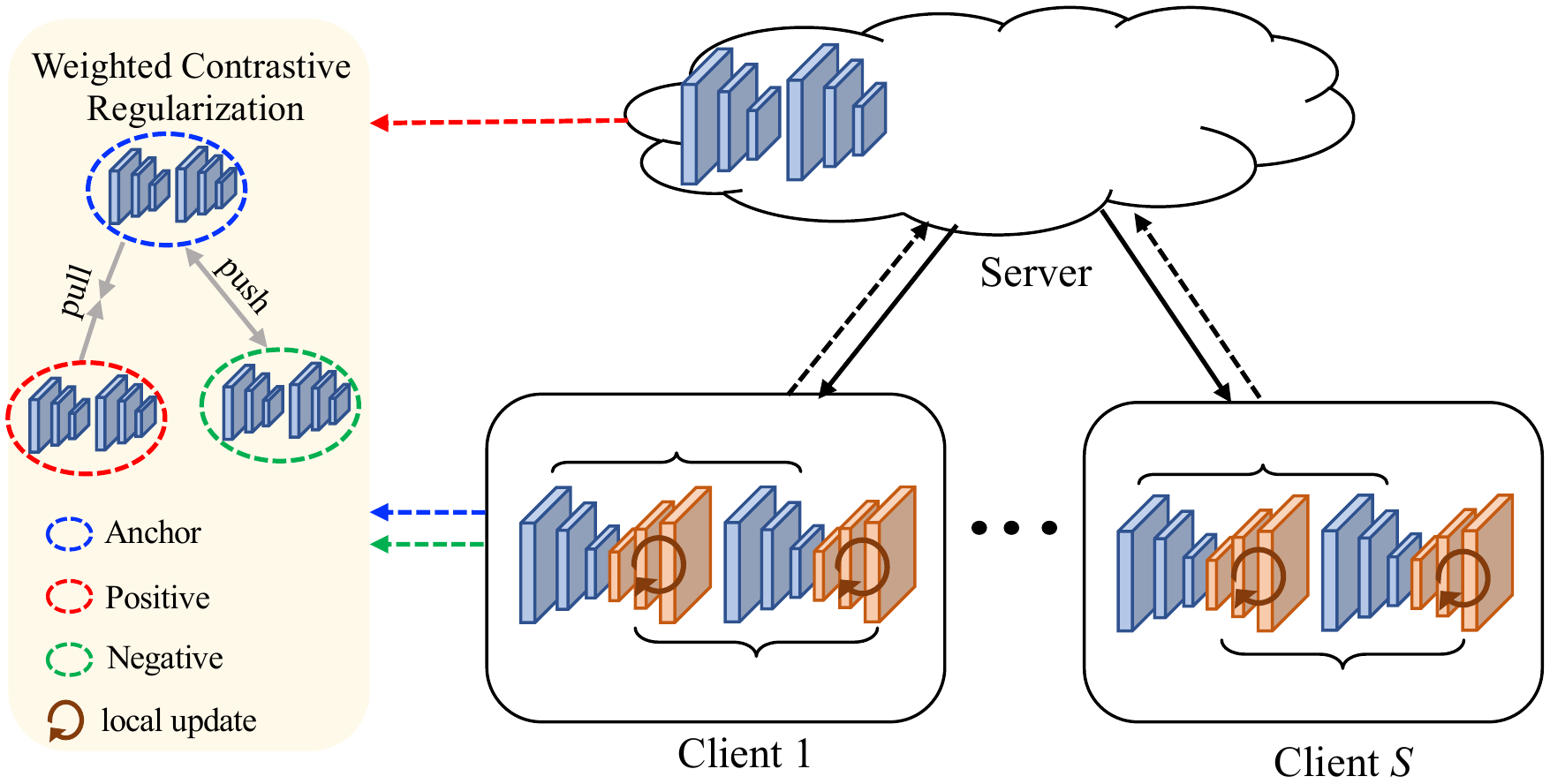}
  \put(-107,109){\scriptsize $\Theta_{{G}_{e}}^{z}\!=\!\frac{1}{S} \sum_s^{S} \Theta_{G_{e}^{s}}^{z-1}$}
  \put(-132,80){\scriptsize $\Theta_{G_{e}^{1}}^{z+1}$} 
  \put(-56,80){\scriptsize $\Theta_{G_{e}^{s}}^{z+1}$} 
  \put(-76,70){\scriptsize $\Theta_{{G}_{e}}^{z}$} 
  \put(-106,70){\scriptsize $\Theta_{{G}_{e}}^{z}$} 
  \put(-176,113){\scriptsize $\Theta_{G_{e}}^{z}$}
  \put(-192,53){\scriptsize $\Theta_{G_{e}^{s}}^{z,t}$} 
  \put(-208,29){\scriptsize $\Theta_{G_{e}^{1,...,S}}^{z-1}$}
  \put(-145,55){\scriptsize $G_{e}$} 
  \put(-135,15){\scriptsize $G_{d}$}
  \put(-54,53){\scriptsize $G_{e}$}
  \put(-39,13){\scriptsize $G_{d}$} 
%   \put(-18,3){\tiny $K$}
  
  \caption{\textbf{Overview of the FedMRI framework.} Instead of averaging all the local client models, a ``globally shared encoder'' \protect\includegraphics[scale=0.09,valign=c]{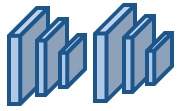} is used to obtain a generalized representation of both frequency and image spaces, and a ``client-specific decoder'' \protect\includegraphics[scale=0.055,valign=c]{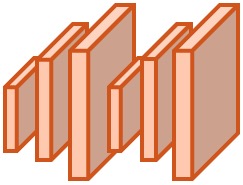} is used to explore unique domain-specific information of the two domains. We apply the weighted contrastive regularization to better pull (\protect\includegraphics[scale=0.01,valign=c]{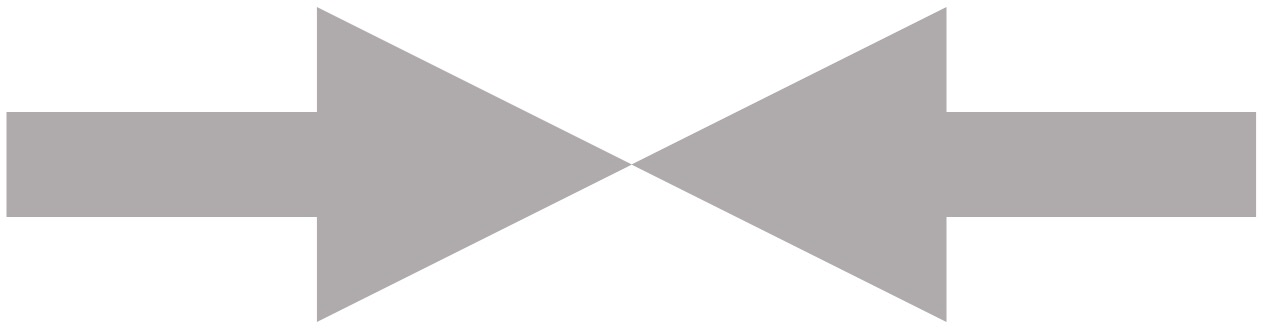}) the positive pairs (\protect\includegraphics[scale=0.016,valign=c]{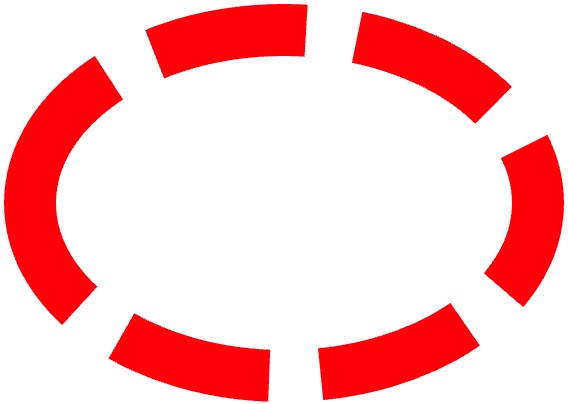}, \protect\includegraphics[scale=0.015,valign=c]{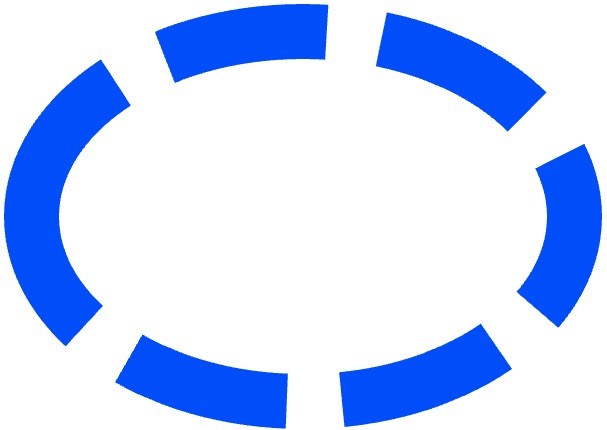}) together and push (\protect\includegraphics[scale=0.01,valign=c]{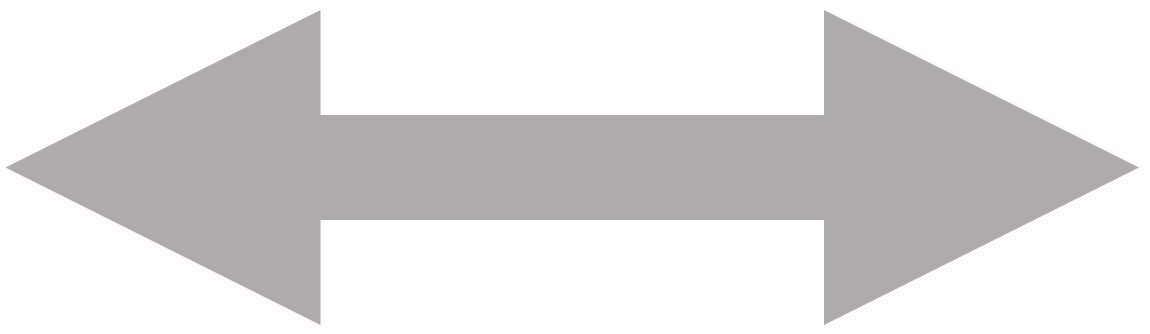}) the negative (\protect\includegraphics[scale=0.016,valign=c]{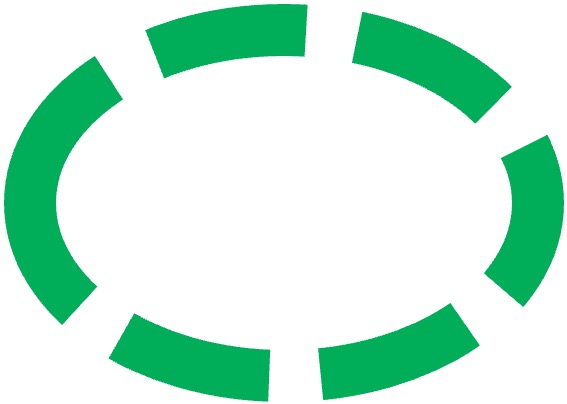}) ones towards the anchor.}
  \label{fig2}
\end{figure}
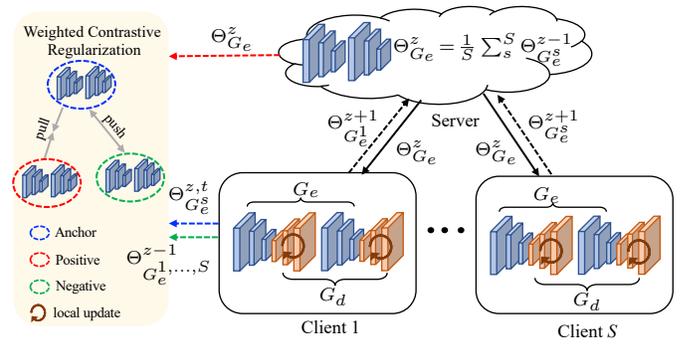

Our network architecture uses a cascade U-Net both in $k$-space and image domains (KI-Net~\cite{eo2018kiki}) as the base framework for clients, where the encoder extracts latent features and the decoder generates images. As shown in~\figref{fig2}, we share the encoder with the central server to extract a global generalized representation of both frequency and image spaces, and the decoder is kept local to explore the unique local depth information of the two domains. To alleviate the bias of local updates, a weighted contrastive regularization term is used to constrain the updating direction of the gradient during the iteration process.

FedMRI aims to find the optimal unique information for each client, while the server and client learn the global generalized representation together (see~\figref{fig2}). To this end, each round of FedMRI communication alternates between client updates and server updates. We consider dividing the reconstruction model $G^s$ into $G_{e}^s=\{G_{e_k}^s,G_{e_i}^s\}$ and $G_{d}^s=\{G_{d_k}^s,G_{d_i}^s\}$ for sharing global and finding unique deep properties in both $k$-space and image domains, respectively. Thus, the objective of Eq.~\eqref{eq:3} can be rewritten as:
\vskip -10pt
\begin{equation}
\begin{split}
\mathcal{L}_{rec}= \frac{1}{S}\sum_{s=1}^{S}\left\| \left(G_{e}(\mathbf{x})\circ G_{d}^s(\mathbf{x}) \right)-\mathbf{y}\right\|_{1},
\end{split}
\end{equation}
where $G_{e}$ and $G_{d}^s$ represent the globally shared encoder and a client-specific decoder, respectively. Note that the global representation $G_{e}$ is jointly learnt by the server and client, while unique domain-specific information is explored by each client. The alternate update rules between the client and server in each round of communication are as follows. Due to the space limitations, in Eq.~\eqref{eq:7} and Eq.~\eqref{eq:8}, we use the $G_{e}^s$ represents $ {G_{e_k}^s,G_{e_i}^s}$, and $G_{d}^s$ represents ${G_{d_k}^s,G_{d_i}^s}$.

% \vspace{0.1cm}
\noindent{\textbf{Client Update Step:}} The client updates the local gradient based on the global representation currently transmitted by the server to find the optimal unique local properties:
\vskip -10pt
\begin{equation}
\Theta_{G_{d}^{s}}^{z,t+1} = \Theta_{G_{d}^{s}}^{z,t}-\eta_{s} \nabla \mathcal{L}\left(\Theta_{G_{e}^{}}^{z}\cup \Theta_{G_{d}^{s}}^{z,t}; \mathbf{x}^{s}\right),
\label{eq:7}
\end{equation}
where $\Theta_{G_{e}}^{z}\cup\Theta_{G_{d}^{s}}^{z,t}=\Theta_{G^{s}}^{z,t}$. The advantage of this update rule is that the number of local updates can be controlled to find the optimal client-specific decoder based on the local data (see section \S\ref{sec:local update}).

\noindent{\textbf{Server Update Step:}} After the local update epoch is complete, the client participates in the global gradient update in the server as follows:
\vskip -10pt
\begin{equation}
\Theta_{G_{e}^{s}}^{z+1} = \Theta_{G_{e}^{s}}^{z}-\eta_{s} \nabla \mathcal{L}\left(\Theta_{G_{e}^{s}}^{z}\cup  \Theta_{G_{d}^{s}}^{z,T}; \mathbf{x}^{s}\right).
\label{eq:8}
\end{equation}
Then, we send $\Theta_{G_{e}^{s}}^{z+1}$ to the server and average them to compute the next global representation $\Theta_{G_{e}}^{z+1}$.

\begin{algorithm}[!t]
    \caption{FedMRI}
    \label{alg:r2p}
    \KwIn{Datasets from $S$ clients: $\mathcal{D}^{1}, \mathcal{D}^{2}, \ldots, \mathcal{D}^{S}$; number of local updates $T$; number of communication rounds $Z$; hyperparameter $\mu$; learning rate for client $s$: $\eta_{s}$;}
    \KwOut{The final global model parameter $\Theta_{G_{e}}$;}

      \For{$z=1,2,...,Z$}
      {

        \For{$s=1,2,...S$}
        {
            receive the global model from the server\;

            \For{$t=1,2,...T$}
            {
                  $\Theta_{G_{d}^{s}}^{z,t+1}$ $\leftarrow$ $\Theta_{G_{d}^{s}}^{z,t}-\eta_{k} \nabla \mathcal{L}_{}$\;
            }
          client locally update the representation and sends it to the server:\\
          $\Theta_{G_{e}^s}^{z+1}$ $\leftarrow$ $\Theta_{G_{e}^{s}}^{z}-\eta_{k} \nabla \mathcal{L}_{}$; 
       
          $\mathcal{L}_{con}$ $\leftarrow$ ContrastiveLoss $\left(\Theta_{G_{e}^{s}}^{z,t} ; \{ \Theta_{G_{e}^{i}}^{z-1},i\!\in\!K\}; \Theta_{G_{e}}^{z} ; \mathbf{x}\right)$;\\
          $\mathcal{L}_{rec}$ $\leftarrow$SupervisedLoss\\$\left(\Theta_{G_{e}}\cup\Theta_{G_{d}^{s}} ;(\mathbf{x}, \mathbf{y})\right)$;\\
          $\mathcal{L}=\mathcal{L}_{rec}+\mu \mathcal{L}_{c o n}$;            
            
        }
    average to get the new representation.\
        
      }

\end{algorithm}

\noindent{\textbf{Weighted Contrastive Regularization:}}\label{sec:we}
Although we have shared the weight $\Theta_{G_{e}^{s}}^{z}$ to the server to find the generalized representation among the clients, there is always an offset between $\Theta_{G_{e}}^{z}$ and $\Theta_{G_{e}^{s}}^{z}$ during iterative optimization, mainly caused by the domain shift during local optimization. To further correct the local updates and provide the model with global recognition ability, we introduce weighted contrastive regularization between the local and global encoders to force ${G_{e}^{s}}$ to learn a stronger generalized representation. Different to construct positive and negative pairs on data, we directly regularize the update direction on the network parameters, which is different from traditional contrastive learning. This enables the gradient updates to be corrected more directly, without relying on a large batch size~\cite{yuan2020contrastive}. Our experimental results also support this, as shown in Table~\ref{t3}.

Suppose the client $k$ performs a local update, it first receives global parameters $\Theta_{G_{e}}^{z}$ from the server and then performs a local iterative update based on these. However, the global parameters from the server always have less bias than the local parameters. Following~\cite{wu2021contrastive}, we use the $\mathcal{L}_1$ distance to measure the difference between the two models. Thus, our weighted contrastive regularization loss can be defined as:
\vskip -10pt
\begin{equation}
\mathcal{L}_{con}=\frac{\left\| \Theta_{G_{e}^{s}}^{z,t}- \Theta_{G_{e}}^{z}\right\|_{1}}{\sum_{i=1}^{S}\left\| \Theta_{G_{e}^{i}}^{z-1}- \Theta_{G_{e}^{s}}^{z,t}\right\|_{1}}.
\end{equation}
Combined with the supervised reconstruction loss, the overall loss of our model can be expressed as:
\vskip -10pt
\begin{equation}
    \begin{aligned}
    \mathcal{L}&=\mathcal{L}_{rec}\left(\Theta_{G_{e}}\cup\Theta_{G_{d}^{s}} ;(\mathbf{x}, \mathbf{y})\right)\\
    &+\mu \mathcal{L}_{c o n}\left(\Theta_{G_{e}^{s}}^{z} ; \{ \Theta_{G_{e}^{i}}^{z-1},i\!\in\!S\} ; \Theta_{G_{e}}^{z} ; \mathbf{x}\right),
   \end{aligned}
\end{equation}%
where $\mu$ is a hyperparameter that controls the weight of the weighted contrastive regularization term.

% The detailed procedure is outlined in Algorithm~\ref{alg:r2p}. 
The detailed procedure is outlined in Algorithm~\ref{alg:r2p}. In each round of communication, the server sends the global model to each client. Each client then performs local gradient updating according to the global representation currently transmitted by the server to acquire its optimal unique information, as shown in Eq.~\eqref{eq:7}. Then, the client participates in the server update according to Eq.~\eqref{eq:8} and corrects the local gradient updates.

\begin{figure}[!t]
\centering
  \includegraphics[width=1\linewidth]{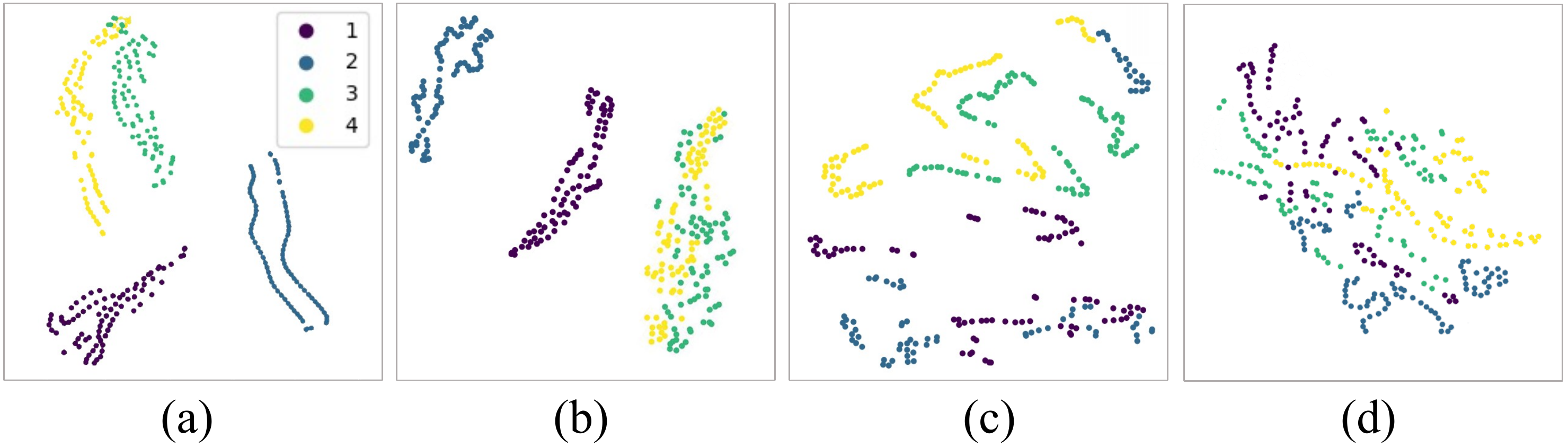}
  \vskip -2pt
  \caption{T-SNE visualizations of latent features from four datasets: \protect\includegraphics[scale=0.01,valign=c]{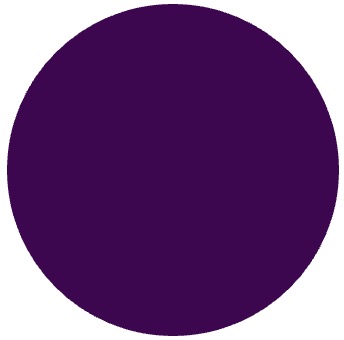} fastMRI~\cite{zbontar2018fastmri},  \protect\includegraphics[scale=0.01,valign=c]{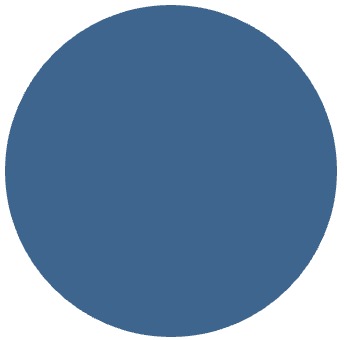}  BraTS~\cite{menze2014multimodal}, \protect\includegraphics[scale=0.01,valign=c]{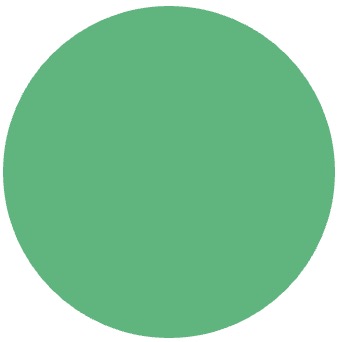} SMS~\cite{feng2021multi}, and \protect\includegraphics[scale=0.01,valign=c]{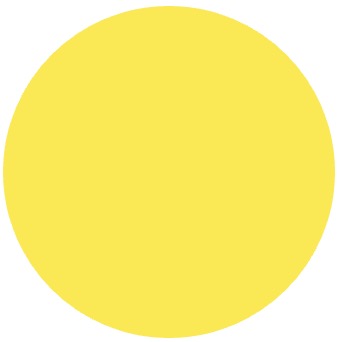} uMR~\cite{feng2021multi}. (a) Distribution of SingleSet, where each client is trained with their local data without FL; (b) Distribution of FedAvg; (c) Distribution of our FedMRI without $\mathcal{L}_{c o n}$; (d) Distribution of our entire FedMRI algorithm.}
  \label{fig1}
%   \vskip -15pt
\end{figure}

\subsection{Comparison with Standard FL}

Our main idea is in line with the domain adaption techniques, where a shared encoder learns a domain-invariant representation to ensure that all inputs are equally suitable for any domain transformation~\cite{hu2018duplex}, while the decoder maintains the domain-special representation. Similarly, in FL for MR image reconstruction, an ideal model should consider both the shared information and client-specific properties in the frequency space and image space respectively. Naturally, we can send the $k$-space domain encoder and image domain encoder of each client back to the server to learn a generalized representation, while the decoders for both domains are kept in local clients to explore the unique client-specific properties.

To further verify our idea, we visualize the T-SNE distributions of latent features in~\figref{fig1}, where (a-d) show the distributions of SingleSet, FedAvg~\cite{mcmahan2017communication}, FedMRI without $\mathcal{L}_{c o n}$, and our entire FedMRI algorithm, respectively. In SingleSet, each client is trained to use their local data without FL. The distribution of points in (a) is clearly differentiated because each dataset has its own biases, while the data in (b), (c) and (d) overlap to varying degrees, as these models benefit from the joint training mechanism of FL. However, on the datasets with large differences in distribution, \eg, fastMRI and BraTS, FedAvg~\cite{mcmahan2017communication} nearly fails (see~\figref{fig1} (b)). Notably, even without $\mathcal{L}_{c o n}$, our method still aligns the latent space distribution across the four different datasets, which demonstrates that sharing a global encoder and keeping a client-specific decoder in both frequency and image spaces can effectively reduce the domain shift problem (see~\figref{fig1} (c)).~\figref{fig1} (d) shows a fully mixed distribution for the latent features of the different clients. This can be attributed to the weighted contrastive regularization, which enables our FedMRI algorithm to effectively correct deviations between the client and server during optimization (see~\figref{fig1} (d)).

%-------------------------------------------------------------------------
\section{Experiments}\label{sec:exp}
In this section, we present the datasets, implementations, and experiments in detail. Our studies \textbf{1)} a comparison in reconstruction accuracy with the current state of the arts under different Scenarios (\texttt{Scenario 1} is a multi-institutional collaboration with local data, \texttt{Scenario 2} is a multi-institutional collaboration with a higher domain shift, and \texttt{Scenario 3} is a multi-institutional collaboration with more local clients); \textbf{2)} an evaluation of communication efficiency and the influence of local updates on the global representation; and \textbf{3)} an ablation study of the key components in our method. 

\subsection{Experimental Setup}

\noindent{\textbf{Implementation Details.}}
We train our method using Pytorch with eight NVIDIA Tesla V$100$ GPUs and $32$GB of memory per card. The batch size and initial learning rate are set to 8 and $1 \times 10^{-4}$, respectively. The models are trained using an RMSProp optimizer, and 50 communication rounds with 10 local epochs each. The hyperparameter $\mu$ is empirically set to 100. We tested and found that our method converges after 50 communication rounds, and achieves the best results when $\mu$=100 (see~\figref{round} and~\figref{epoch}).

% \vspace{0.1cm}
\noindent{\textbf{Baselines.}}
We compare against various state-of-the-art models to demonstrate the effectiveness of our method. These include: (1) SingleSet, in which each client is trained using their local data without FL; (2) FedAvg~\cite{mcmahan2017communication}, which trains a global model by averaging the parameters from all participating clients; (3) FL-MRCM~\cite{guo2021multi}, which removes the domain shift during multi-institutional MR imaging collaboration by aligning the latent features between the source and target clients; (4) GD-GD~\cite{arivazhagan2019federated}, which explores a personalized layer after the base part; (5) LG-FedAvg~\cite{liang2020think}, which attempts to learn the local heads and a global network body, as an opposite to our approach; (6) FedBN~\cite{li2021fedbn}, which alleviates the domain shift by applying local batch normalization; and (7) FedProx~\cite{li2018federated}, which adds a proximal term to the objective function of the local model. (8) Transfer-Site, where the model is transferred across different sites in a random order.

Following~\cite{guo2021multi}, we adopt U-Net~\cite{zbontar2018fastmri} as the reconstruction network of each client for all the baselines. To further evaluate the effectiveness of our mechanism in both $k$-space and image space, we also test most representative baselines, \eg, FedAvg, FedProx, and GD-GD, that use the KI-Net~\cite{eo2018kiki} (two cascade U-Net) as the reconstruction network of each client. For all classification FL methods, we changed their supervised loss to the $\mathcal{L}_1$ loss for MR reconstruction. For FL-MRCM, we use its default loss, \eg, $\mathcal{L}_1$ loss. To handle data imbalances across different institutions, the server can distribute the global model according to the client data proportion when aggregating the local model.

% \vspace{0.1cm}
\noindent{\textbf{Datasets.}}
We employ four different datasets with different distributions as the clients in the FL setting. Each dataset is divided with a ratio of 7:3 for local \texttt{train/test}.
\begin{itemize} 
	\item \textbf{fastMRI}~\cite{zbontar2018fastmri}: $2,\!134$ brain subjects with $16$ T1 and $16$ T2 weighted slices each are used in our experiments. 
	
	\item \textbf{BraTS}~\cite{menze2014multimodal}: $385$ brain subjects with $80$ T1 weighted and $80$ T2 weighted slices each are adopt.
	
	\item \textbf{SMS}~\cite{feng2021multi}: $155$ subjects are collected by a 3T Siemens Magnetom Skyra system, with $20$ T1 and $20$ T2 weighted slices each.	
	
	\item \textbf{uMR}~\cite{feng2021multi}: $50$ subjects are collected by a United Imaging Healthcare uMR $790$ scanner, with $20$ T1 weighted and $20$ T2 weighted slices each.  
% 	\vspace{-4pt}
\end{itemize}

\begin{table*}[!t]
	\centering
	\caption{\small \textbf{Quantitative comparison} of state-of-the-art FL methods under \texttt{Scenario 1}, with \textit{1D uniform} sampling and \textit{3$\times$} acceleration. $^\dagger$ indicates the baseline methods that use the KI-Net as the reconstruction network of each client. $P<$ 0.001 was considered as a statistically significant level. See \S\ref{sec:cs1} for more details.}
	\small
	\resizebox{1\textwidth}{!}{
		\setlength\tabcolsep{2pt}
		\renewcommand\arraystretch{1.2}
		\begin{tabular}{r||ccc|ccc|ccc|ccc|ccc}
			\hline\thickhline
			\rowcolor{mygray}
		    &\multicolumn{3}{c|}{\textbf{fastMRI}~\cite{zbontar2018fastmri}} &\multicolumn{3}{c|}{\textbf{BraTS}~\cite{menze2014multimodal}}&\multicolumn{3}{c|}{\textbf{SMS}~\cite{feng2021multi}}&\multicolumn{3}{c|}{\textbf{uMR}~\cite{feng2021multi}}&\multicolumn{3}{c}{\textbf{Average}}  \\ 
			\rowcolor{mygray}
			\multirow{-2}{*}{Method}&PSNR$\uparrow$&SSIM$\uparrow$&$P$ values &PSNR$\uparrow$&SSIM$\uparrow$&$P$ values&PSNR$\uparrow$&SSIM$\uparrow$&$P$ values&PSNR$\uparrow$&SSIM$\uparrow$&$P$ values&PSNR$\uparrow$&SSIM$\uparrow$ &$P$ values\\\hline\hline
			Transfer-Site&19.24&0.298&$<0.001$/$<0.001$&23.47&0.203&$<0.001$/$<0.001$&35.53&0.967&$<0.001$/$<0.001$&36.72&0.940&$<0.001$/$<0.001$&28.81&0.602&$<0.001$/$<0.001$ \\
            SingleSet&36.88&0.936&$<0.001$/$<0.001$&32.74&0.911&$<0.001$/$<0.001$&25.20&0.758&$<0.001$/$<0.001$&23.50&0.683&$<0.001$/$<0.001$&29.58&0.822&$<0.001$/$<0.001$ \\
            FedAvg~\cite{mcmahan2017communication}&36.56&0.933&$<0.001$/$<0.001$&32.90&0.908&$<0.001$/$<0.001$&25.25&0.722&$<0.001$/$<0.001$&21.09&0.646&$<0.001$/$<0.001$&28.95&0.802&$<0.001$/$<0.001$\\
            FedBN~\cite{li2021fedbn}&29.40&0.853&$<0.001$/$<0.001$&28.85&0.713&$<0.001$/$<0.001$&29.44&0.855&$<0.001$/$<0.001$&30.73&0.879&$<0.001$/$<0.001$&29.61&0.825&$<0.001$/$<0.001$\\
            FL-MRCM~\cite{guo2021multi}&30.05&0.736&$<0.001$/$<0.001$&30.50&0.780&$<0.001$/$<0.001$&30.66&0.859&$<0.001$/$<0.001$&29.38&0.888&$<0.001$/$<0.001$&30.04&0.816&$<0.001$/$<0.001$\\
            FedProx~\cite{li2018federated}&36.13&0.902&$<0.001$/$<0.001$&31.16&0.813&$<0.001$/$<0.001$&29.02&0.887&$<0.001$/$<0.001$&29.54&0.889&$<0.001$/$<0.001$&31.46&0.873&$<0.001$/$<0.001$\\\hline
            LG-FedAvg~\cite{liang2020think}&37.04&0.940&$<0.001$/$<0.001$&32.52&0.899&$<0.001$/$<0.001$&30.00&0.834&$<0.001$/$<0.001$&28.31&0.861&$<0.001$/$<0.001$&31.97&0.884&$<0.001$/$<0.001$\\
            GD-GD~\cite{arivazhagan2019federated}&38.05&0.959&$<0.001$/$<0.001$&32.89&0.939&$<0.001$/$<0.001$&34.79&0.939&$<0.001$/$<0.001$&35.90&0.950&$<0.001$/$<0.001$&35.41&0.947&$<0.001$/$<0.001$\\\hline
            FedAvg$^\dagger$~\cite{liang2020think}&35.42&0.926&$<0.001$/$<0.001$&33.12&0.910&$<0.001$/$<0.001$&26.10&0.725&$<0.001$/$<0.001$&21.56&0.640&$<0.001$/$<0.001$&29.05&0.800&$<0.001$/$<0.001$\\
            FedProx$^\dagger$~\cite{li2018federated}&36.22&0.912&$<0.001$/$<0.001$&31.25&0.810&$<0.001$/$<0.001$&29.00&0.890&$<0.001$/$<0.001$&29.67&0.903&$<0.001$/$<0.001$&31.54&0.878&$<0.001$/$<0.001$\\
            GD-GD$^\dagger$~\cite{arivazhagan2019federated}&38.26&0.958&$<0.001$/$<0.001$&32.91&0.940&$<0.001$/$<0.001$&35.00&0.952&$<0.001$/$<0.001$&36.11&0.961&$<0.001$/$<0.001$&35.57&0.955&$<0.001$/$<0.001$\\\hline
            \textbf{FedMRI}&{\textbf{39.21}}&{\textbf{0.970}}&$<0.001$/$<0.001$&{\textbf{33.93}}&{\textbf{0.940}}&$<0.001$/$<0.001$&{\textbf{39.76}}&{\textbf{0.974}}&$<0.001$/$<0.001$&{\textbf{38.20}}&{\textbf{0.969}}&$<0.001$/$<0.001$&{\textbf{37.59}} &{\textbf{0.964}}&$<0.001$/$<0.001$\\\hline
		\end{tabular}
	}
	\captionsetup{font=small}
	\label{t1}
\end{table*}

\begin{table*}[!t]
	\centering
	\caption{\small \textbf{Quantitative comparison} of state-of-the-art FL methods under \texttt{Scenario 2},  with different undersampling patterns and various acceleration rates (\eg, \textit{1D uniform 3$\times$}, \textit{1D Cartesian 5$\times$}, \textit{2D radial 4$\times$}, and \textit{2D random 6$\times$}). $^\dagger$ indicates the baseline methods that use the KI-Net as the reconstruction network of each client. $P<$ 0.001 was considered as a statistically significant level. See \S\ref{sec:cs2} for more details.}
	\small
	\resizebox{1\textwidth}{!}{
		\setlength\tabcolsep{2pt}
		\renewcommand\arraystretch{1.2}
		\begin{tabular}{r||ccc|ccc|ccc|ccc|ccc}
			\hline\thickhline
			\rowcolor{mygray}
		    &\multicolumn{3}{c|}{\textbf{fastMRI}~\cite{zbontar2018fastmri}} &\multicolumn{3}{c|}{\textbf{BraTS}~\cite{menze2014multimodal}}&\multicolumn{3}{c|}{\textbf{SMS}~\cite{feng2021multi}}&\multicolumn{3}{c|}{\textbf{uMR}~\cite{feng2021multi}}&\multicolumn{3}{c}{\multirow{3}{*}{\textbf{Average}}}  \\
		    \rowcolor{mygray}
		    {Method}&\multicolumn{3}{c|}{\textit{1D uniform 3$\times$}}&\multicolumn{3}{c|}{\textit{1D Cartesian 5$\times$}}&\multicolumn{3}{c|}{\textit{2D radial 4$\times$}}&\multicolumn{3}{c|}{\textit{2D random 6$\times$}}&\multicolumn{3}{c}{\multirow{-2}{*}{\textbf{Average}}}\\
			\rowcolor{mygray}
			\multirow{-2}{*}{}&PSNR$\uparrow$&SSIM$\uparrow$&$P$ values &PSNR$\uparrow$&SSIM$\uparrow$&$P$ values&PSNR$\uparrow$&SSIM$\uparrow$&$P$ values&PSNR$\uparrow$&SSIM$\uparrow$&$P$ values&PSNR$\uparrow$&SSIM$\uparrow$ &$P$ values\\\hline\hline
			Transfer-Site&18.76&0.283&$<0.001$/$<0.001$&22.03&0.137&$<0.001$/$<0.001$&33.20&0.957&$<0.001$/$<0.001$&36.22&0.938&$<0.001$/$<0.001$&27.53&0.579&$<0.001$/$<0.001$ \\
            SingleSet&35.26&0.925&$<0.001$/$<0.001$&29.66&0.869&$<0.001$/$<0.001$&24.49&0.771&$<0.001$/$<0.001$&24.25&0.661&$<0.001$/$<0.001$&28.41&0.806&$<0.001$/$<0.001$\\
            FedAvg~\cite{mcmahan2017communication}&35.53&0.920&$<0.001$/$<0.001$&29.82&0.872&$<0.001$/$<0.001$&23.95&0.734&$<0.001$/$<0.001$&24.19&0.675&$<0.001$/$<0.001$&28.37&0.800&$<0.001$/$<0.001$\\
            FedBN~\cite{li2021fedbn}&31.28&0.787&$<0.001$/$<0.001$&29.95&0.885&$<0.001$/$<0.001$&26.87&0.817&$<0.001$/$<0.001$&26.73&0.790&$<0.001$/$<0.001$&28.71&0.820&$<0.001$/$<0.001$\\
            FL-MRCM~\cite{guo2021multi}&30.16&0.772&$<0.001$/$<0.001$&28.04&0.760&$<0.001$/$<0.001$&28.48&0.857&$<0.001$/$<0.001$&29.33&0.864&$<0.001$/$<0.001$&29.00&0.813&$<0.001$/$<0.001$\\
            FedProx~\cite{li2018federated}&35.29&0.920&$<0.001$/$<0.001$&29.82&0.881&$<0.001$/$<0.001$&28.80&0.895&$<0.001$/$<0.001$&29.28&0.902&$<0.001$/$<0.001$&30.79&0.900&$<0.001$/$<0.001$\\\hline
            LG-FedAvg~\cite{liang2020think}&36.75&0.952&$<0.001$/$<0.001$&29.91&0.873&$<0.001$/$<0.001$&32.56&0.923&$<0.001$/$<0.001$&33.35&0.923&$<0.001$/$<0.001$&33.14&0.918&$<0.001$/$<0.001$\\
            GD-GD~\cite{arivazhagan2019federated}&38.00&0.959&$<0.001$/$<0.001$&30.07&0.902&$<0.001$/$<0.001$&37.02&0.970&$<0.001$/$<0.001$&37.28&0.959&$<0.001$/$<0.001$&35.59&0.948&$<0.001$/$<0.001$\\\hline
            FedAvg$^\dagger$~\cite{liang2020think}&35.61&0.924&$<0.001$/$<0.001$&29.92&0.888&$<0.001$/$<0.001$&24.15&0.740&$<0.001$/$<0.001$&24.28&0.680&$<0.001$/$<0.001$&28.49&0.808&$<0.001$/$<0.001$\\
            FedProx$^\dagger$~\cite{li2018federated}&35.41&0.933&$<0.001$/$<0.001$&29.96&0.885&$<0.001$/$<0.001$&28.89&0.900&$<0.001$/$<0.001$&29.30&0.914&$<0.001$/$<0.001$&30.89&0.908&$<0.001$/$<0.001$\\
            GD-GD$^\dagger$~\cite{arivazhagan2019federated}&38.12&0.961&$<0.001$/$<0.001$&30.15&0.922&$<0.001$/$<0.001$&37.25&0.978&$<0.001$/$<0.001$&37.40&0.960&$<0.001$/$<0.001$&35.73&0.955&$<0.001$/$<0.001$\\\hline
            \textbf{FedMRI}&{\textbf{39.99}}&{\textbf{0.986}}&$<0.001$/$<0.001$&{\textbf{32.35}}&{\textbf{0.934}}&$<0.001$/$<0.001$&{\textbf{39.57}}&{\textbf{0.991}}&$<0.001$/$<0.001$&{\textbf{38.14}}&{\textbf{0.970}}&$<0.001$/$<0.001$&{\textbf{37.51}} &{\textbf{0.970}}&$<0.001$/$<0.001$\\\hline
		\end{tabular}
	}
	\captionsetup{font=small}

	\label{t2}
\end{table*}

\subsection{Experimental Results}

\noindent\textbf{Accuracy Comparison.}
To verify the effectiveness of our FL approach, FedMRI, in multi-institutional collaboration, we compare it with various state-of-the-art FL algorithms. For fair comparison, we retrain them on MR data and report their optimal results. For FedProx, we tune its hyperparameter $\mu$ within $\{0.001,0.01,0.1,1\}$ and report the best result at $0.01$. For other methods, we use their default parameters and retrain them on the same datasets. All the compared methods are trained over $70$ communication rounds with $10$ local epochs each, and they all converge at round $70$ (see section \S\ref{sec:communicate}). Table~\ref{t1} shows the reconstruction results of all methods under \texttt{Scenario 1}, which employs \textit{1D uniform} sampling with \textit{3$\times$} acceleration. We use the paired Student’s t-test to evaluate the significant difference between the two methods. $P<$ 0.001 was considered as a statistically significant level. Our FedMRI is superior to all the other methods in both PSNR and SSIM. In particular, on small datasets, such as \textbf{SMS} and \textbf{uMR}, our method improves the PSNR results of FedAvg from $25.25$ dB and $21.09$ dB to $\textbf{39.76}$ dB and $\textbf{38.2}$ dB, respectively. However, Transfer-site has the lowest results because it cannot solve the non-IID problem. The model can only concentrate on the current client, and the previous model will be forgotten after transfer to other clients. More importantly, even compared with the methods that also used KI-Net as the client model, our FedMRI is still significantly higher than them. This demonstrates that the client update mechanism (see Eq.~\eqref{eq:7}.) of our FedMRI can effectively remove the domain shift in the non-IID datasets. The distributed training on data from multiple institutions without compromising privacy can solve the dependence of traditional deep learning on the collection of big data.\label{sec:cs1}

% \vspace{0.1cm}
\noindent\textbf{Robustness to a Higher Domain-Shift Setting.}
To further evaluate the robustness of our method, we record the results of each client under different sampling patterns and accelerations in Table~\ref{t2}. Considering that various hospital scanners use different scanners, we employ different undersampling patterns, \eg, \textit{1D uniform 3$\times$}, \textit{1D Cartesian 5$\times$}, \textit{2D radial 4$\times$}, and \textit{2D random 6$\times$}, to set up \texttt{Scenario 2}. As expected, our method achieves better results under a higher domain shift. In contrast, the FedAvg and MRI multi-institutional collaboration method, FL-MRCM, performs worse under this scenario. Compared with other methods, FedMRI is most robust because it can balance the learning of local and global representations, ensuring that the client learns unique client-specific properties. \label{sec:cs2}

 \begin{figure*}[!t]
	\begin{center}
		\includegraphics[width=0.9\linewidth]{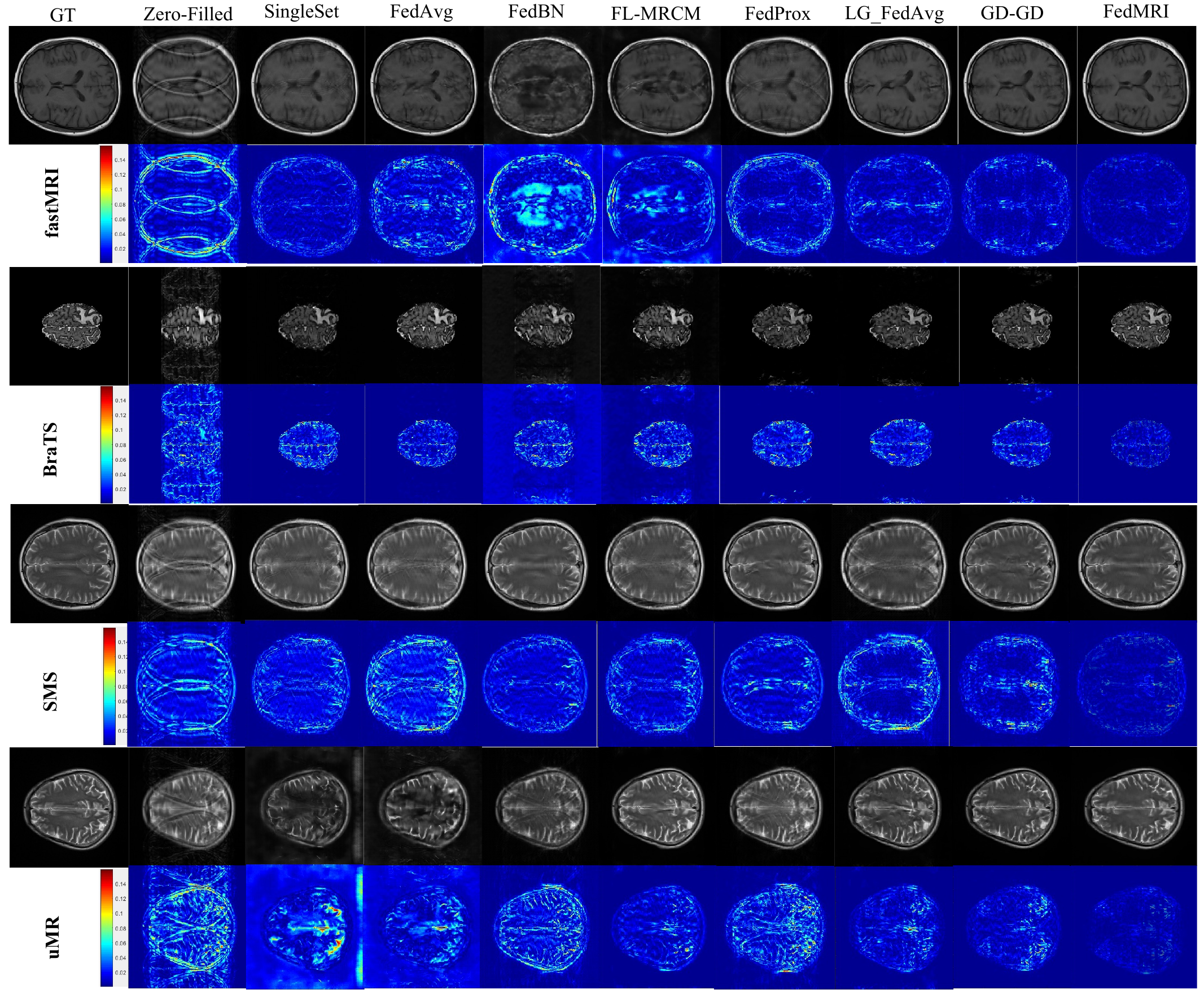}
	\end{center}
% 	\vspace{-10pt}
	\captionsetup{font=small}
	\caption{\small\textbf{Qualitative comparison} of different methods in terms of reconstruction quality and error maps under \texttt{Scenario 1}.}
	\label{fig:a1}%\vspace{-0.4cm}
\end{figure*}

\begin{figure*}[!t]
\centering
  \includegraphics[width=0.90\textwidth]{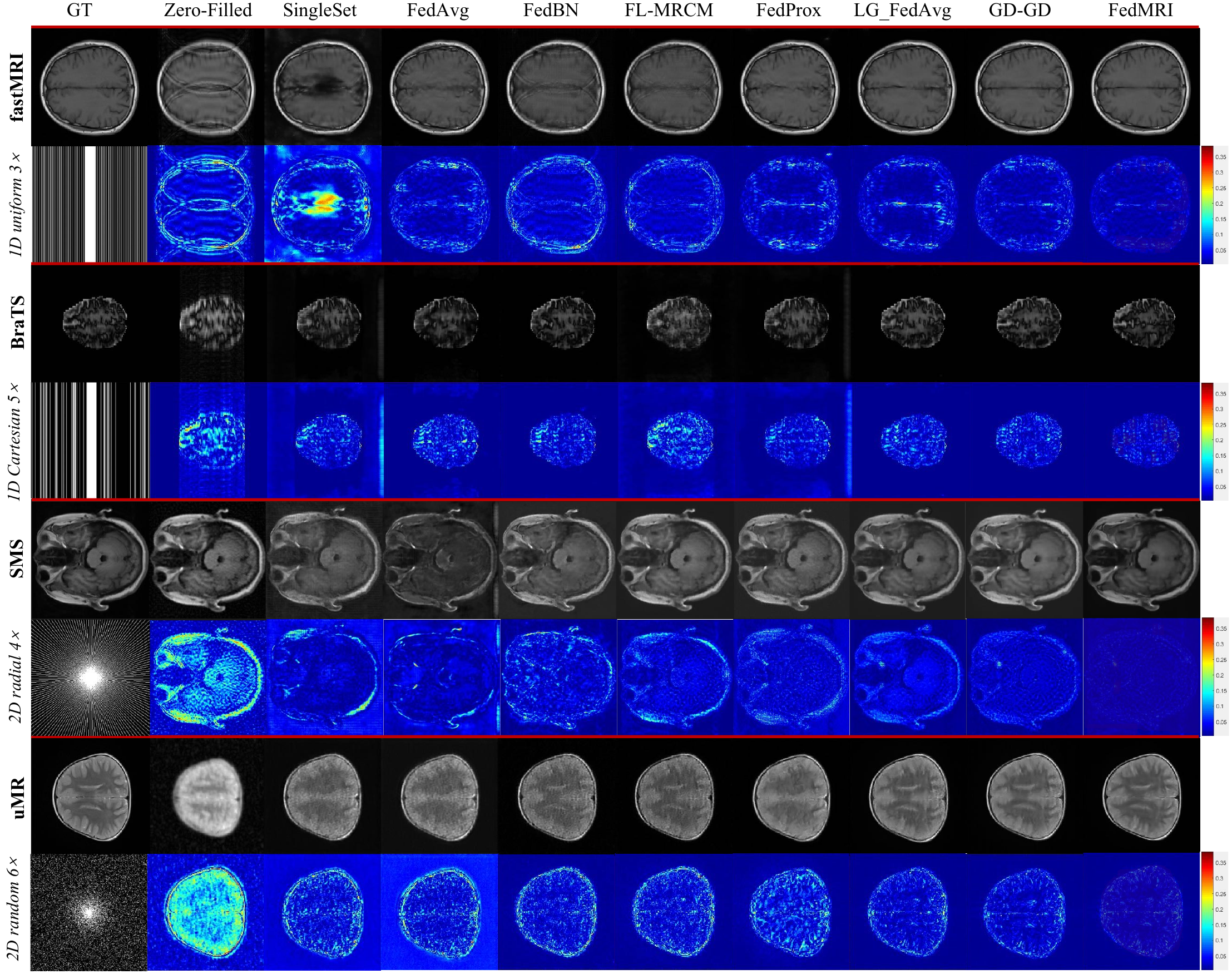}
  \caption{\textbf{Qualitative comparison} of different methods in terms of reconstruction quality and error maps under \texttt{Scenario 2}.}
  \label{error2}
\end{figure*}

In addition, it is worth noting that our method also outperforms the personalized FL methods, \eg, GD-GD~\cite{arivazhagan2019federated} and LG-FedAvg~\cite{liang2020think}, in both \texttt{Scenario 1} and \texttt{2}. Since GD-GD only keeps the last layer as the personalized layer after the base part, which damages the strong unique local properties, there is almost no improvement in the higher domain shift setting. LG-FedAvg shares a global decoder/network body and learns a local encoder/head, which is opposite to our approach, resulting in a poor generalized representation. The results support our previous suggestion that an ideal personalized mechanism must train shared information before the client-specific property. Therefore, the most effective personalized mechanism for MR image reconstruction is to use latent features to obtain a generalized representation first and then preserve the unique domain-specific properties in each client.

To further evaluate the robustness of our method, we further divided each dataset into two clients according to different sampling patterns to construct eight clients. The detailed settings and average results of each method are shown in Table~\ref{txx3}. As can be seen form this Table, when the number of clients changed, FedMRI still maintained outstanding reconstruction results.

\begin{table*}[!t]
	\centering
	\caption{\small \textbf{Quantitative comparison} of state-of-the-art FL methods under \texttt{Scenario 3},  with different undersampling patterns and various acceleration rates (\eg, \textit{1D uniform 3$\times$}, \textit{2D radial 2$\times$}, \textit{1D Cartesian 5$\times$}, \textit{2D random 4$\times$}, \textit{2D radial 4$\times$}, \textit{1D uniform 5$\times$}, \textit{2D random 6$\times$}, and \textit{1D Cartesian 3$\times$}). $^\dagger$ indicates the baseline methods that use the KI-Net as the reconstruction network of each client.}
	\small
	\resizebox{0.95\textwidth}{!}{
		\setlength\tabcolsep{7pt}
		\renewcommand\arraystretch{1.0}
		\begin{tabular}{r||cc|cc|cc|cc|cc|cc}
			\hline\thickhline
			
			\rowcolor{mygray}
		{Metrics}	&PSNR$\uparrow$&SSIM$\uparrow$&PSNR$\uparrow$&SSIM$\uparrow$&PSNR$\uparrow$&SSIM$\uparrow$&PSNR$\uparrow$&SSIM$\uparrow$&PSNR$\uparrow$&SSIM$\uparrow$&PSNR$\uparrow$&SSIM$\uparrow$\\\hline\hline

			\rowcolor{mygray}
		   {Methods}  &\multicolumn{2}{c|}{Transfer-Site}&\multicolumn{2}{c|}{SingleSet} &\multicolumn{2}{c|}{FedAvg~\cite{mcmahan2017communication}}&\multicolumn{2}{c|}{FedBN~\cite{li2021fedbn}}&\multicolumn{2}{c|}{FL-MRCM~\cite{guo2021multi}}&\multicolumn{2}{c}{FedProx~\cite{li2018federated}}\\

   {Average}&27.23&0.562&28.46&0.800&28.28&0.799&28.70&0.830&28.86&0.830&30.82&0.912\\
            
    			\rowcolor{mygray}
		   {Methods}  &\multicolumn{2}{c|}{LG-FedAvg~\cite{liang2020think}} &\multicolumn{2}{c|}{GD-GD~\cite{arivazhagan2019federated}}&\multicolumn{2}{c|}{FedAvg$^\dagger$~\cite{liang2020think}}&\multicolumn{2}{c|}{FedProx$^\dagger$~\cite{li2018federated}}&\multicolumn{2}{c|}{GD-GD$^\dagger$~\cite{arivazhagan2019federated}}&\multicolumn{2}{c}{\textbf{FedMRI}}\\
    {Average}&33.22&0.920&35.45&0.937&28.50&0.818&30.92&0.912&35.80&0.960&\textbf{37.49}&\textbf{0.971}\\\hline
            
		\end{tabular}
	}
	\captionsetup{font=small}

	\label{txx3}
\end{table*}

\begin{table*}[!t]
	\centering
	\caption{\small \textbf{Ablation study of the key components} in our method, where $G_{e_k}^s$, $G_{d_k}^s$, $G_{e_i}^s$ and $G_{d_i}^s$ are the shared part of the FL model, and $\mathcal{L}_{c o n}^f$ and $\mathcal{L}_{c o n}^w$ represent the contrastive regularization based on features and weights, respectively. See \S\ref{sec:ab1} for more details.}
	\small
	\resizebox{0.95\textwidth}{!}{
		\setlength\tabcolsep{7pt}
		\renewcommand\arraystretch{1.0}
		\begin{tabular}{r||cccccc|cc|cc|cc|cc|cc}
			\hline\thickhline
			\rowcolor{mygray}
            &&&&&&&\multicolumn{2}{c|}{\textbf{fastMRI}~\cite{zbontar2018fastmri}}&\multicolumn{2}{c|}{\textbf{BraTS}~\cite{menze2014multimodal}}&\multicolumn{2}{c|}{\textbf{SMS}~\cite{feng2021multi}}&\multicolumn{2}{c|}{\textbf{uMR}~\cite{feng2021multi}}&\multicolumn{2}{c}{\multirow{2}{*}{\textbf{}}}\\
            \rowcolor{mygray}
		    Methods&$G_{e_k}^s$&$G_{d_k}^s$&$G_{e_i}^s$&$G_{d_i}^s$&$\mathcal{L}_{c o n}^f$&$\mathcal{L}_{c o n}^w$&\multicolumn{2}{c|}{\textit{1D uniform 3$\times$}}&\multicolumn{2}{c|}{\textit{1D Cartesian 5$\times$}}&\multicolumn{2}{c|}{\textit{2D radial 4$\times$}}&\multicolumn{2}{c|}{\textit{2D random 6$\times$}}&\multicolumn{2}{c}{\multirow{-2}{*}{\textbf{Average}}}\\
			\rowcolor{mygray}
            &&&&&&&PSNR&SSIM&PSNR&SSIM&PSNR&SSIM&PSNR&SSIM&PSNR&SSIM\\\hline\hline
            
            $\mathcal{M}_1$ &\Checkmark&\Checkmark&\Checkmark&\Checkmark&\XSolidBrush&\XSolidBrush&35.53&0.920&29.82&0.872&23.95&0.734&24.19&0.675&28.37&0.800\\
            $\mathcal{M}_2$ &\Checkmark&\Checkmark&\Checkmark&\XSolidBrush&\XSolidBrush&\XSolidBrush&38.42 &0.962 &30.24 &0.915 &37.36 &0.968 &37.44 &0.959 &35.87 &0.951 \\
            $\mathcal{M}_3$ &\Checkmark&\Checkmark&\Checkmark&\XSolidBrush&\Checkmark&\XSolidBrush&36.29 &0.938 &29.07 &0.805 &34.84 &0.876 &34.08 &0.841 &33.57 &0.865 \\
            $\mathcal{M}_4$&\XSolidBrush&\XSolidBrush&\Checkmark&\XSolidBrush&\XSolidBrush&\Checkmark&38.25 &0.945 &30.19 &0.928 &38.56 &0.982 &38.00 &0.959 &36.25 &0.953 \\\hline
            $~\textbf{\texttt{Ours}}$&\Checkmark&\XSolidBrush&\Checkmark&\XSolidBrush&\XSolidBrush&\Checkmark&{\textbf{39.99}}&{\textbf{0.986}}&{\textbf{32.35}}&{\textbf{0.934}}&{\textbf{39.57}}&{\textbf{0.991}}&{\textbf{38.14}}&{\textbf{0.970}}&{\textbf{37.51}} &{\textbf{0.970}}\\ \hline
		\end{tabular}
	}
	\captionsetup{font=small}

	\label{t3}

\end{table*}

To qualitatively evaluate our method, we visualize the reconstructed images and corresponding error maps of all the compared methods for \texttt{Scenario 1} in~\figref{fig:a1} and \texttt{Scenario 2} in~\figref{error2}. The less texture in the error map, the better the reconstruction. As can be seen from the figure, our method reconstructs the best-quality images and significantly reduces errors. The results of \texttt{Scenario 2} are consistent with those of \texttt{Scenario 1}, indicating that \textit{our FedMRI’s reconstructed results are robust under the presence of a domain shift.}

\begin{figure}[!t]
\centering
  \includegraphics[width=1\linewidth]{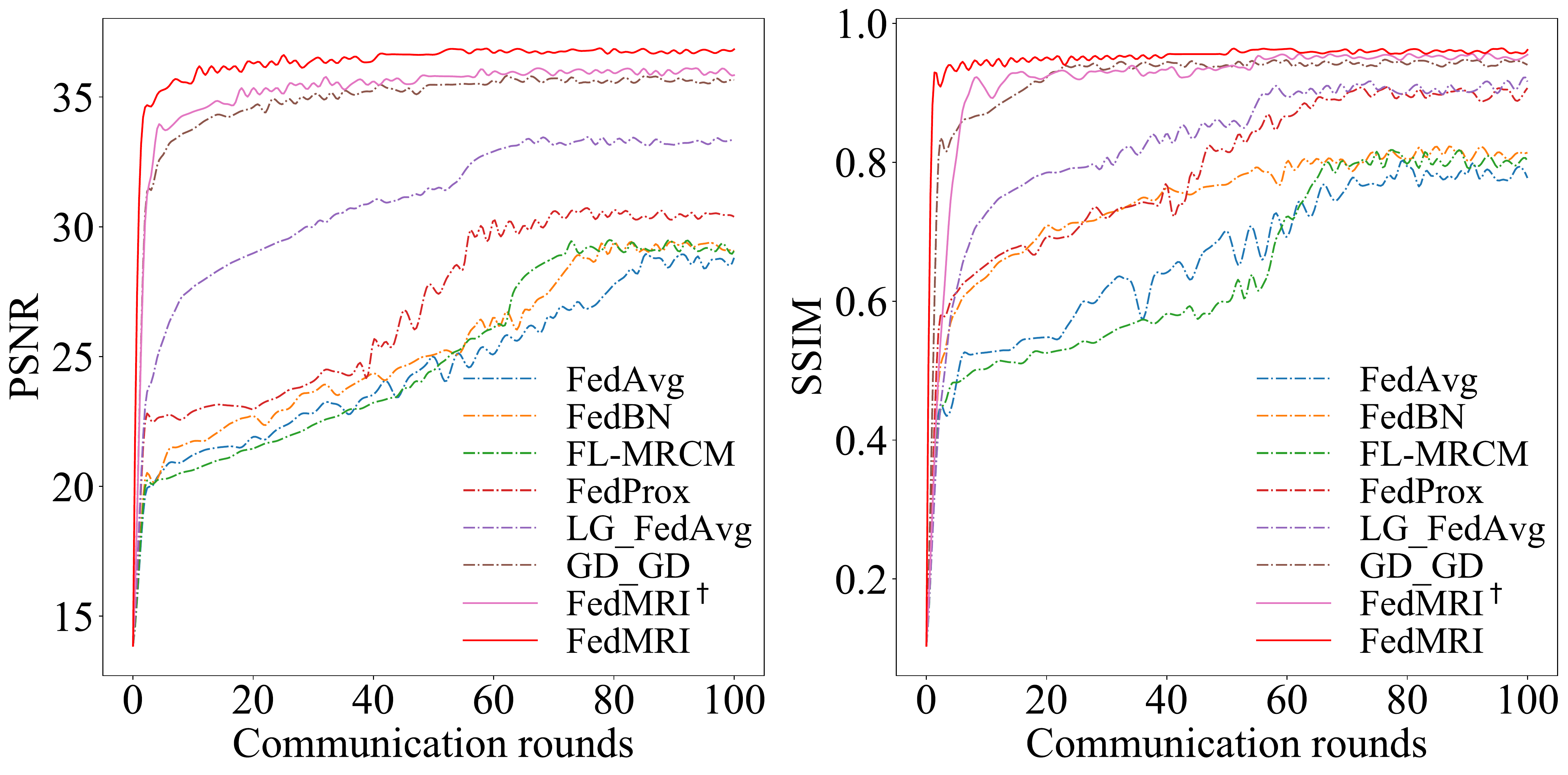}
  \caption{\textbf{Analysis of communication efficiency} in terms of PSNR and SSIM.}
  \label{round}
\end{figure}

% \vspace{0.1cm}
\noindent\textbf{Communication Efficiency.} In~\figref{round}, we compare the accuracy of all methods in each round to demonstrate the superior communication efficiency of our method, where FedMRI$^\dagger$ represents our model without weighted contrastive regularization. Note that the local epochs are set to $10$ for all the compared methods. As can be seen from the figure, FedMRI$^\dagger$ achieves stability in the fifth round. The other methods improve most in the first five rounds, but do not achieve stability until after more than $50$ rounds. With the help of weighted contrastive regularization, our FedMRI converges much faster than FedMRI$^\dagger$ because it corrects the deviation between the client and server during optimization. As we have mentioned that weighted contrastive regularization can boost the convergence, the results in this figure confirm our previous conjecture. This means that our method has the highest communication efficiency, which is one of the important issues in FL~\cite{konevcny2016federated}.\label{sec:communicate}

\begin{figure}[!t]
\centering
  \includegraphics[width=1\linewidth]{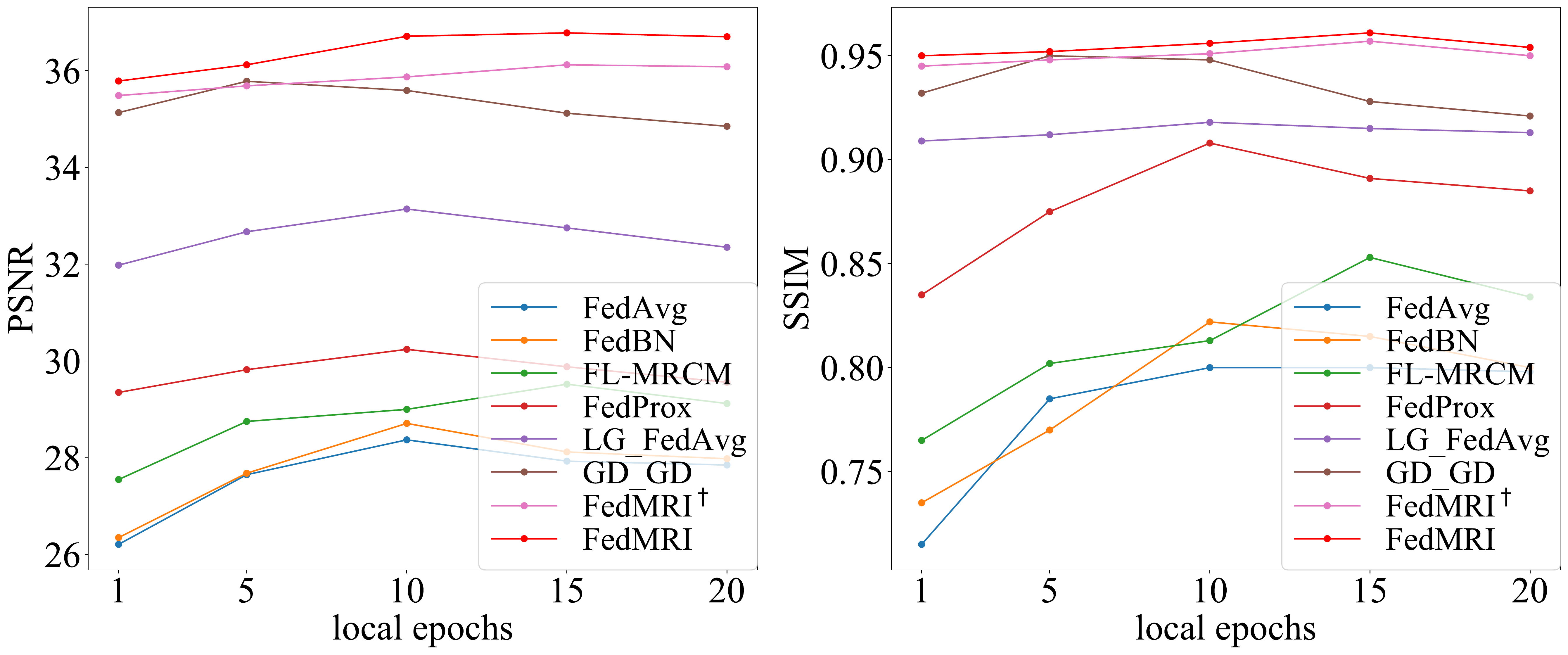}
  \caption{\textbf{Analysis of different numbers of local epochs} in terms of PSNR and SSIM.}
  \label{epoch} 
  \vskip -5pt
\end{figure}

% \vspace{0.1cm}
\noindent\textbf{Benefit of Local Updates.}\label{sec:local update} The domain shift problem between clients causes the server to deviate from the global optimal solution when averaging. As mentioned in \S\ref{sec:intro}, considering both the shared generalized representation and client-specific properties helps mitigate this problem. In~\figref{epoch}, we explore whether more local epochs can enhance the generalized representation to address the bias from the global optimal solution. From this figure, we observe that 
increasing the number of local epochs improves both FedMRI$^\dagger$ and FedMRI significantly compared to other FL methods. Further, when the number of local epochs is too large, the reconstruction accuracy of other methods decreases. In contrast, for our method, more local epochs can alleviate the divergence from the global optimal solution without damaging local computations.

\subsection{Ablation Study}
In this section, we first investigate the crucial components in our method, \eg, the shared part of the local client ($G_{e_k}^s$, $G_{d_k}^s$, $G_{e_i}^s$ and $G_{d_i}^s$), and weighted contrastive regularization. For comparison, we build two baselines, $\mathcal{M}_1$ and $\mathcal{M}_2$, which represent our model without the partially shared network and weighted contrastive regularization respectively. As shown in Table~\ref{t3}, we conduct tests under \texttt{Scenario 2}. $\mathcal{M}_1$ performs the worst, indicating that partially sharing the representation makes it possible to train on different clients with different domains. Since the weighted contrastive regularization term corrects the offsets of local client updates, the generalized representation learned by $\mathcal{M}_2$ is not optimal when this term is lost. In addition, we also investigate the effect of using traditional contrastive learning (see $\mathcal{M}_3$), which adopts the positive and negative feature pairs, as mentioned in~\cite{li2021model}. As can be seen, $\mathcal{M}_3$ is far inferior to directly imposing contrastive regularization on the parameters, which can directly correct the gradient updates and does not depend on a large batch size. Our FedMRI obtains the best results thanks to its two crucial parts, proving its strong ability in removing the domain shift. To evaluate the effectiveness of our partial shared mechanism in the hybrid domain, we record the results of $\mathcal{M}_4$ that only performed on the image domain. As can be seen from these results, our method still obtains high accuracy on the single domain demonstrating that our specificity-preserving mechanism can help the MR image reconstruction in multi-institution joint training.

\label{sec:ab1}

\begin{figure}[!t]
\centering
  \includegraphics[width=1\linewidth]{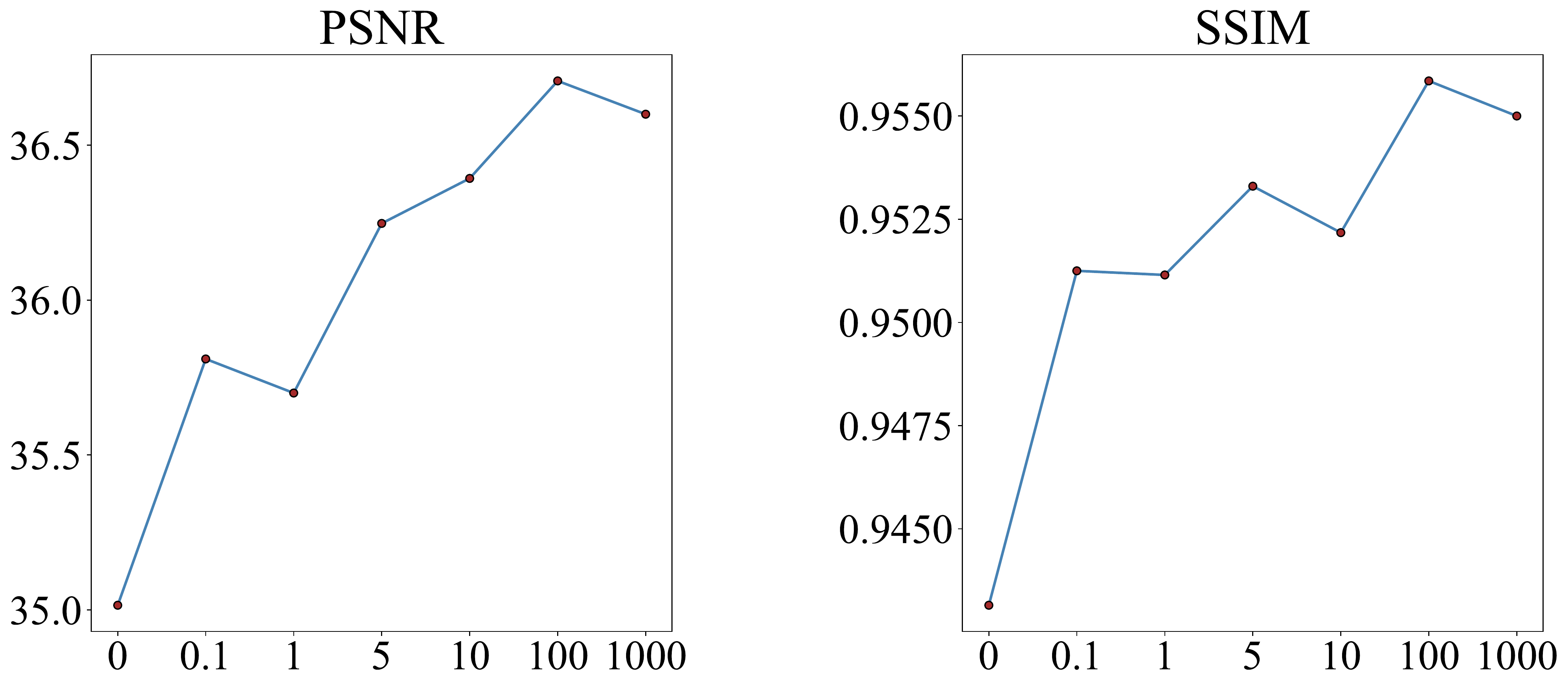}
  \caption{\textbf{Analysis of weighted contrastive regularization} hyperparameter $\mu$ in terms of PSNR and SSIM.}
  \vskip -10pt
  \label{mu}
\end{figure}

Finally, we investigate the influence of the weighted contrastive regularization hyperparameter $\mu$ in our method. \figref{mu} shows the PSNR and SSIM values from $\{0, 1, 10, 100, 100\}$. From the figure, we observe that the reconstruction accuracy increases with increasing $\mu$. Setting $\mu=100$ yields the best results, with the weighted contrastive regularization leading to substantial performance gains in PSNR (\ie, $35.87$ dB$\rightarrow$ $\textbf{37.51}$ dB). When $\mu$ is greater than 100, the performance degrades due to an overly large weight, which affects the unique information learning of the local client. \label{sec:ab2}

\begin{table}[!t]
	\centering
	\caption{\small\textbf{Comparison of the average training times per round \texttt{Scenario 2}.} FedMRI$^\dagger$ represents our FedMRI without $\mathcal{L}_{c o n}$ See \S\ref{sec:cost} for more details.}
	\small
	\resizebox{0.4\textwidth}{!}{
		\setlength\tabcolsep{32pt}
		\renewcommand\arraystretch{1.0}
		\begin{tabular}{r||c}
			\hline\thickhline
			\rowcolor{mygray}
			Methods & Times \\ \hline\hline
			FedAvg~\cite{mcmahan2017communication}&80min    \\
			FedBN~\cite{li2021fedbn}&80min    \\
			FL-MRCM~\cite{guo2021multi}&145min     \\
			FedProx~\cite{li2018federated}&83min   \\
			LG-FedAvg~\cite{liang2020think}&75min   \\
			GD-GD~\cite{arivazhagan2019federated}&76min    \\
			\hline\hline
			FedMRI$^\dagger$ &73min   \\
			FedMRI &84min   \\
			\hline
		\end{tabular}
	}
	\captionsetup{font=small}

	\label{table:ap2} 
\end{table}

\begin{table*}[!t]
	\centering
	\caption{\small \textbf{Comparison of different sharing proportions} under \texttt{Scenario 2}, where $~\textbf{\texttt{Ours}}_{1}$, $~\textbf{\texttt{Ours}}_{2}$, and $~\textbf{\texttt{Ours}}_{3}$ represent our method with the entire local model shared, with the entire local model except the last layer shared, and with only the first downsampling layer shared. $P<$ 0.001 was considered as a statistically significant level. See \S\ref{sec:sharebili} for more details.}
	\small
	\resizebox{1\textwidth}{!}{
		\setlength\tabcolsep{2pt}
		\renewcommand\arraystretch{1.2}
		\begin{tabular}{r||ccc|ccc|ccc|ccc|ccc}
			\hline\thickhline
			\rowcolor{mygray}
            &\multicolumn{3}{c|}{\textbf{fastMRI}~\cite{zbontar2018fastmri}}&\multicolumn{3}{c|}{\textbf{BraTS}~\cite{menze2014multimodal}}&\multicolumn{3}{c|}{\textbf{SMS}~\cite{feng2021multi}}&\multicolumn{3}{c|}{\textbf{uMR}~\cite{feng2021multi}}&\multicolumn{3}{c}{\multirow{3}{*}{\textbf{Average}}}\\
            \rowcolor{mygray}
		    Methods&\multicolumn{3}{c|}{\textit{1D uniform 3$\times$}}&\multicolumn{3}{c|}{\textit{1D Cartesian 5$\times$}}&\multicolumn{3}{c|}{\textit{2D radial 4$\times$}}&\multicolumn{3}{c|}{\textit{2D random 6$\times$}}&\multicolumn{3}{c}{\multirow{-2}{*}{\textbf{Average}}}\\
			\rowcolor{mygray}
            &PSNR$\uparrow$&SSIM$\uparrow$&$P$ values &PSNR$\uparrow$&SSIM$\uparrow$&$P$ values&PSNR$\uparrow$&SSIM$\uparrow$&$P$ values&PSNR$\uparrow$&SSIM$\uparrow$&$P$ values&PSNR$\uparrow$&SSIM$\uparrow$ &$P$ values\\\hline\hline
            $~\textbf{\texttt{Ours}}_{1}$&35.26&0.925&$<0.001$/$<0.001$&29.66&0.869&$<0.001$/$<0.001$&24.49&0.771&$<0.001$/$<0.001$&24.25&0.661&$<0.001$/$<0.001$&28.41&0.806&$<0.001$/$<0.001$\\
            $~\textbf{\texttt{Ours}}_{2}$ &38.00&0.959&$<0.001$/$<0.001$&30.07&0.902&$<0.001$/$<0.001$&37.02&0.970&$<0.001$/$<0.001$&37.28&0.959&$<0.001$/$<0.001$&35.59&0.948&$<0.001$/$<0.001$\\
            $~\textbf{\texttt{Ours}}_{3}$&36.50&0.935&$<0.001$/$<0.001$&29.25&0.905&$<0.001$/$<0.001$&31.63&0.941&$<0.001$/$<0.001$&31.22&0.931&$<0.001$/$<0.001$&32.15 &0.928&$<0.001$/$<0.001$\\
            $~\textbf{\texttt{Ours}}$
            &{\textbf{39.99}}&{\textbf{0.986}}&$<0.001$/$<0.001$&{\textbf{32.35}}&{\textbf{0.934}}&$<0.001$/$<0.001$&{\textbf{39.57}}&{\textbf{0.991}}&$<0.001$/$<0.001$&{\textbf{38.14}}&{\textbf{0.970}}&$<0.001$/$<0.001$&{\textbf{37.51}} &{\textbf{0.970}}&$<0.001$/$<0.001$\\

 \hline
		\end{tabular}

	}

	\label{table:ap1}
% 	\vspace{-6pt}
\end{table*}

\begin{table*}[!t]
	\centering
	\caption{\small \textbf{Comparison of different loss designs} for the weighted contrastive regularization under \texttt{Scenario 2}, where NT-Xent-Contrastive, $\mathcal{L}_2$-Contrastive, and $\mathcal{L}_1$-Contrastive represent our weighted contrastive regularization term with the NT-Xent, $\mathcal{L}_2$, and $\mathcal{L}_1$ norm, respectively, and \textit{w/o}-Contrastive represents our model without the weighted contrastive regularization term. $P<$ 0.001 was considered as a statistically significant level. See \S\ref{sec:loss} for more details.}
	\small
	\resizebox{1\textwidth}{!}{
		\setlength\tabcolsep{2pt}
		\renewcommand\arraystretch{1.2}
		\begin{tabular}{r||ccc|ccc|ccc|ccc|ccc}
			\hline\thickhline
			\rowcolor{mygray}
            &\multicolumn{3}{c|}{\textbf{fastMRI}~\cite{zbontar2018fastmri}}&\multicolumn{3}{c|}{\textbf{BraTS}~\cite{menze2014multimodal}}&\multicolumn{3}{c|}{\textbf{SMS}~\cite{feng2021multi}}&\multicolumn{3}{c|}{\textbf{uMR}~\cite{feng2021multi}}&\multicolumn{3}{c}{\multirow{3}{*}{\textbf{Average}}}\\
            \rowcolor{mygray}
		    Methods&\multicolumn{3}{c|}{\textit{1D uniform 3$\times$}}&\multicolumn{3}{c|}{\textit{1D Cartesian 5$\times$}}&\multicolumn{3}{c|}{\textit{2D radial 4$\times$}}&\multicolumn{3}{c|}{\textit{2D random 6$\times$}}&\multicolumn{3}{c}{\multirow{-2}{*}{\textbf{Average}}}\\
			\rowcolor{mygray}
            &PSNR$\uparrow$&SSIM$\uparrow$&$P$ values &PSNR$\uparrow$&SSIM$\uparrow$&$P$ values&PSNR$\uparrow$&SSIM$\uparrow$&$P$ values&PSNR$\uparrow$&SSIM$\uparrow$&$P$ values&PSNR$\uparrow$&SSIM$\uparrow$ &$P$ values\\\hline\hline
            
            \textit{w/o}-Contrastive&38.42 &0.962&$<0.001$/$<0.001$ &30.24 &0.915&$<0.001$/$<0.001$ &37.36 &0.968&$<0.001$/$<0.001$ &37.44 &0.959&$<0.001$/$<0.001$ &35.87 &0.951&$<0.001$/$<0.001$\\
            NT-Xent-Contrastive &39.00&0.965&$<0.001$/$<0.001$&30.87&0.914&$<0.001$/$<0.001$&36.37&0.960&$<0.001$/$<0.001$&36.69&0.951&$<0.001$/$<0.001$&35.73&0.948&$<0.001$/$<0.001$\\
            $\mathcal{L}_2$-Contrastive &38.98&0.966&$<0.001$/$<0.001$&30.69&0.913&$<0.001$/$<0.001$&37.21&0.970&$<0.001$/$<0.001$&37.02&0.960&$<0.001$/$<0.001$&35.98&0.952&$<0.001$/$<0.001$ \\
            $\mathcal{L}_1$-Contrastive&{\textbf{39.99}}&{\textbf{0.986}}&$<0.001$/$<0.001$&{\textbf{32.35}}&{\textbf{0.934}}&$<0.001$/$<0.001$&{\textbf{39.57}}&{\textbf{0.991}}&$<0.001$/$<0.001$&{\textbf{38.14}}&{\textbf{0.970}}&$<0.001$/$<0.001$&{\textbf{37.51}} &{\textbf{0.970}}&$<0.001$/$<0.001$\\

    \hline
    
	\end{tabular}
	}
	\captionsetup{font=small}
	\label{table:ap3} 
\end{table*}

\subsection{Discussion of Sharing Proportion}\label{sec:sharebili}

Since \texttt{Scenario 2} has a higher domain-shift, we use this scenario to compare different sharing proportions. Table~\ref{table:ap1} records the various models with different sharing proportions, where $~\textbf{\texttt{Ours}}_{1}$, $~\textbf{\texttt{Ours}}_{2}$, and $~\textbf{\texttt{Ours}}_{3}$ represent our method with the entire local model shared, with the entire local model except the last layer shared, and with only the first downsampling layer shared. We observe that $~\textbf{\texttt{Ours}}_{1}$ yields the worst results because sharing the entire local model prevents client-specific information from being preserved. From the results of $~\textbf{\texttt{Ours}}_{2}$ and $~\textbf{\texttt{Ours}}_{3}$, we find that the higher the proportion of sharing, the less effective the reconstruction. However, $~\textbf{\texttt{Ours}}$ achieves the best performance in terms of PSNR and SSIM, suggesting that \textit {designing an effective specificity-preserving FL mechanism for MR image reconstruction is a promising direction.}

% \subsection{Additional Qualitative Results}\label{sec:errormap}

\subsection{Computational Cost}\label{sec:cost}
To evaluate the speed of our algorithm, we record the training time of all comparison methods on an NVIDIA Tesla V100 GPU under \texttt{Scenario 2} in Table~\ref{table:ap2}. As can be seen from this table, the training times of our FedMRI$^\dagger$, LG-FedAvg, and GD-GD are shorter than those of other methods, because they only send a portion of the local model to the server, making them much smaller than the other methods. The training of FedMRI is slower than these partially shared methods because of the introduction of the weighted contrastive regularization term. However, compared with FedAvg~\cite{mcmahan2017communication}, FedBN~\cite{li2021fedbn}, and FedProx~\cite{li2018federated}, our FedMRI has almost no difference in time consumption, but the reconstruction accuracy is definitely improves. Thus, \textit{the computational time of our method is acceptable considering the superior results obtained.} In contrast, the training time of FL-MRCM is twice as long as that of FedMRI$^\dagger$, mainly due to the repeated adversarial training between the target and source clients.

\subsection{Comparison to Other Losses}\label{sec:loss}
We further study the effectiveness of different types of losses on our weighted contrastive regularization term, including $\mathcal{L}_2$ and NT-Xent Loss~\cite{chen2020simple}. Table~\ref{table:ap3} summarizes the reconstruction results of different losses in terms of PSNR and SSIM, where NT-Xent-Contrastive, $\mathcal{L}_2$-Contrastive, and $\mathcal{L}_1$-Contrastive represent our weighted contrastive regularization term with the NT-Xent, $\mathcal{L}_2$, and $\mathcal{L}_1$ norm, respectively, and \textit{w/o}-Contrastive represents our model without the weighted contrastive regularization term. It is worth noting from this figure that our weighted contrastive regularization using the $\mathcal{L}_1$ loss achieves the best performance, outperforming the NT-Xent loss by $\textbf{1.78}$ dB, and the $\mathcal{L}_2$ loss by $\textbf{1.53}$ dB. Compared with \textit{w/o}-Contrastive, we observe that simply using the NT-Xent loss cannot improve the accuracy, and even reduces the reconstruction accuracy of some clients.

Different from most of the current contrastive learning methods, our weighted contrastive regularization explores model weights to replace features and find positive and negative pairs, aiming to correct the direction of local gradient optimization through weighted contrastive regularization. Thus, the local model can be updated while maintaining the performance of the global model. Compared with feature-based contrastive learning, the weighted contrastive regularization method gets rid of the dependence on large batch sizes and yields better reconstruction results (see the ablation study).

\section{Conclusion}
In this work, we focus on the domain shift problem in MR image reconstruction using multi-institutional data in a privacy-protected manner. To this end, we propose a specificity-preserving federated learning for MR image reconstruction (FedMRI). By employing a globally shared encoder to learn a generalized representation and a client-specific decoder to explore unique domain-specific properties in both $k$-space domain and image domain, our proposed FedMRI can obtain the most effective global representation and simultaneously preserve the domain characteristics of client-side. Further, the weighted contrastive regularization term corrects the bias of local clients during each iteration, forcing the shared encoder to learn the global representation. Extensive experiments on multiple datasets from different domains demonstrate that our method can achieve MR image reconstruction joint multi-institutional client. Given the need for privacy protection in medical data, we expect further innovations in MR image reconstruction by federated learning.

\textbf{Acknowledgement}
We would like to express our gratitude to Prof. Ender Konukoglu of ETH Zurich for his valuable comments on this work.

{\small
\bibliographystyle{IEEEtran}
\bibliography{ref}
}

\end{document}